\documentclass[notitlepage,twocolumn,showpacs,preprintnumbers,amsmath,amssymb,superscriptaddress, prl,floatfix]{revtex4-1}
\usepackage{bm}
\usepackage{dcolumn}
\usepackage{tabularx}
\usepackage{bm}
\usepackage{amsmath}
\usepackage{color}
\usepackage{here}
\usepackage{graphicx}
\usepackage{setspace}
\usepackage{url}

\begin{document}

\preprint{Preprint}

\title{Probing the Ginzburg-Landau Potential for Lasers Using Higher-order Photon Correlations}

\author{Naotomo Takemura}
\affiliation{NTT Nanophotonics Center, NTT Corp., 3-1, Morinosato Wakamiya Atsugi, Kanagawa 243-0198, Japan}
\affiliation{NTT Basic Research Laboratories, NTT Corp., 3-1, Morinosato Wakamiya Atsugi, Kanagawa 243-0198, Japan}
\author{Masato Takiguchi}
\affiliation{NTT Nanophotonics Center, NTT Corp., 3-1, Morinosato Wakamiya Atsugi, Kanagawa 243-0198, Japan}
\affiliation{NTT Basic Research Laboratories, NTT Corp., 3-1, Morinosato Wakamiya Atsugi, Kanagawa 243-0198, Japan}
\author{Masaya Notomi}
\email[E-mail: ]{masaya.notomi.mn@hco.ntt.co.jp}
\affiliation{NTT Nanophotonics Center, NTT Corp., 3-1, Morinosato Wakamiya Atsugi, Kanagawa 243-0198, Japan}
\affiliation{NTT Basic Research Laboratories, NTT Corp., 3-1, Morinosato Wakamiya Atsugi, Kanagawa 243-0198, Japan}

\date{\today}

\begin{abstract}
Lasing transition  is known to be analogous to the second-order phase transition. Furthermore, for some cases, it is possible to define the Ginzburg-Landau (GL) potential, and the GL theory predicts the photon statistical properties of lasers. However, the GL potential for lasers is surprising, because lasers are operating far from equilibrium. In this paper, we theoretically examine the validity of the GL theory for lasers in terms of various parameters, particularly, the ratio between photon and carrier lifetimes. For this purpose, we use stochastic rate equations and higher-order photon correlation functions. With higher-order photon correlation measurements, we can check whether or not laser dynamics are described by the GL theory. We demonstrate that, for low-$\beta$ lasers, the GL theory is applicable even when the photon lifetime is comparable to the carrier lifetime and that photon-carrier relaxation oscillation is the fundamental origin of the breakdown of the GL theory, which can be understood in the framework of center manifold reduction. 
\end{abstract}

\maketitle
\section{1. Introduction}
An analogy between lasing and a second-order phase transition was pointed out by Haken \cite{Haken2012} and Scully \cite{Scully1999} in the early 70s. Furthermore, they discovered that the photon statistics of lasers can be described by the Ginzburg-Landau (GL) potential. The existence of a thermodynamic potential such as the GL potential for lasers is very surprising because lasing occurs far from equilibrium. Since lasing is a non-equilibrium steady-state, there is a probability current associated with pumping and dissipation. Thus, for lasers, the detailed balance condition is clearly violated, and a thermodynamic potential does not exist in a strict sense. In fact, the success of the laser-phase-transition analogy lies in the fact it was established in gas lasers such as He:Ne lasers \cite{Scully1999,Haken2012,Louisell1973}. For gas lasers, since the photon lifetime is much longer than the other atomic lifetimes (class-A lasers), the atomic degree of freedom is safely eliminated by so-called adiabatic elimination. The slow field dynamics of gas lasers such as He:Ne lasers are the motions in the GL potential. Importantly, the slow field dynamics are equivalent to the equilibrium dynamics in the GL potential. Furthermore, with the inclusion of field noises, the GL theory is known to describe the photon statistical properties of class-A lasers. On the other hand, for widely used semiconductor lasers, the existence of a thermodynamic potential is highly questionable. Since the carrier lifetime is longer than the photon lifetime (class-B lasers), the adiabatic elimination of the carrier degree of freedom is not applicable, and thus their equilibrium description is not possible. In fact, pioneering theoretical \cite{Oppo1985,Paoli1988,Ogawa1989,Ogawa1990,Hofmann2000,Hofmann2000a,Lien2002} and experimental studies \cite{Lien2001,Wang2017} have shown evidence that the photon statistics of semiconductor (class-B) lasers are qualitatively different from that of gas (class-A) lasers described by the GL potential. In particular, in Appendix D, we briefly discuss how our results can be connected to the Toda oscillator approach to class-B lasers. 
 
In this paper, extending these pioneering studies, we re-examine to what extent the photon statistics of semiconductor lasers are described by the GL potential in terms of various physical parameters. Note that we consider lasers with a sufficiently small spontaneous emission coupling coefficient $\beta\ll1$ and focus on the ratio between photon and carrier lifetimes. We performed stochastic numerical simulations using the Langevin equations. To check whether or not the photon statistics are described by the GL potential, instead of calculating the probability distribution function with the corresponding Fokker-Planck equations, we introduce a novel method using normalized higher-order photon (intensity) correlations $g^{(q)}$ \cite{Young1994}, which was originally proposed in the context of the quark-gluon plasma transition \cite{Hwa1992,Hwa1993}. Furthermore, experimentally, the advantage of using $g^{(q)}$ is their loss independence. 

Using the numerical simulations of the Langevin equations and higher-order correlations, we explore the parameter regime where the GL-like theory is valid. As pointed out in previous studies \cite{Paoli1988,Ogawa1989,Lien2001,Wang2017}, when the carrier lifetime is much shorter than the photon lifetime, the GL theory is not valid. Meanwhile, for intermediate cases, for instance, when their lifetimes are the same, the photon statistics can be described by the GL theory. This is surprising because, when they are the same, adiabatic elimination is impossible in a conventional sense. In fact, we found that the applicability of the GL theory is much wider than normally expected. To understand these findings, we employ the center manifold reduction theory, which is an extension of adiabatic elimination \cite{Haken1977}. The wide applicability of the GL theory can be interpreted as a demonstration of the ``slaving principle" proposed by Haken \cite{Haken1977}, which states that, around a bifurcation point, the slow dynamics of a system are governed by order parameters (in our case, the cavity field). These results may motivate a revision of laser classification, and could also provide a laser design principle to optimize photon statistical properties. 

\section{2. Classification of lasers}
First, we briefly summarize the Arecchi's classification of lasers, which employs three decay rates \cite{Arecchi1984,Arecchi2012}: photon $\gamma_c$, polarization $\gamma_\perp$, and population inversion (carrier) decay rates $\gamma_\|$. Lasers are classified as follows:

(i)  Class-A lasers ($\gamma_\perp,\gamma_\|\gg\gamma_c$): When the photon decay rate from a cavity is much smaller than the other decay rates, the adiabatic elimination of both polarization and carrier degrees of freedom from the Maxwell-Bloch equation is possible. The class-A laser dynamics are described solely with the cavity photons  (field). Since the approximated photon equation of motion represents equilibrium  dynamics (satisfy the detailed-balance condition), the photon statistics are obtained analytically with the master equation \cite{Scully1967,Takemura2021} or the Fokker-Planck equation approach \cite{Lax1967,Risken1996,Haken2012}. Furthermore, since the analogy between the lasing transition and second-order phase transition is transparent, the GL theory of lasers has been established for class-A lasers \cite{Graham1970,DeGiorgio1970}.

(ii) Class-B lasers ($\gamma_\perp\gg\gamma_c\gtrsim\gamma_\|$): Since the polarization decay (dephasing) is much faster than the other dynamics, the adiabatic elimination of the polarization from the Maxwell-Bloch equations results in the Statz-deMars rate equations for the photon number (field intensity) $I$ and carrier number (population inversion) $N$ \cite{Rice1994}, which are the commonly used rate equations for semiconductor lasers. Since the photon and carrier dynamics of class-B lasers are non-equilibrium (violate the detailed-balance condition), their photon statistics cannot be obtained analytically. Additionally, class-B  lasers exhibit photon-carrier damped oscillation around the lasing threshold, which is known as photon-carrier relaxation oscillation \cite{Takemura2012,Wang2015}.

Importantly, in this paper, we assume that the dephasing rate $\gamma_\perp$ is always much larger than the other decay rates and that the polarization degree of freedom is adiabatically eliminated. Thus, we do not consider class-C lasers, where all three decay rates are on the same order.

\section{3. Theory}
Now, we introduce the Ginzburg-Landau (GL) theory of lasers, which was developed in Refs \cite{Graham1970,DeGiorgio1970}. Conventionally, the Langevin equations are given, and then their steady-state probability distribution is discussed. However, in this section, we take the opposite approach.  We first discuss a type of probability distribution given by the GL potential and its photon statistical properties. Then, we introduce the Langevin equations that give  the GL probability distribution. 
\begin{figure}
\includegraphics[width=0.35\textwidth]{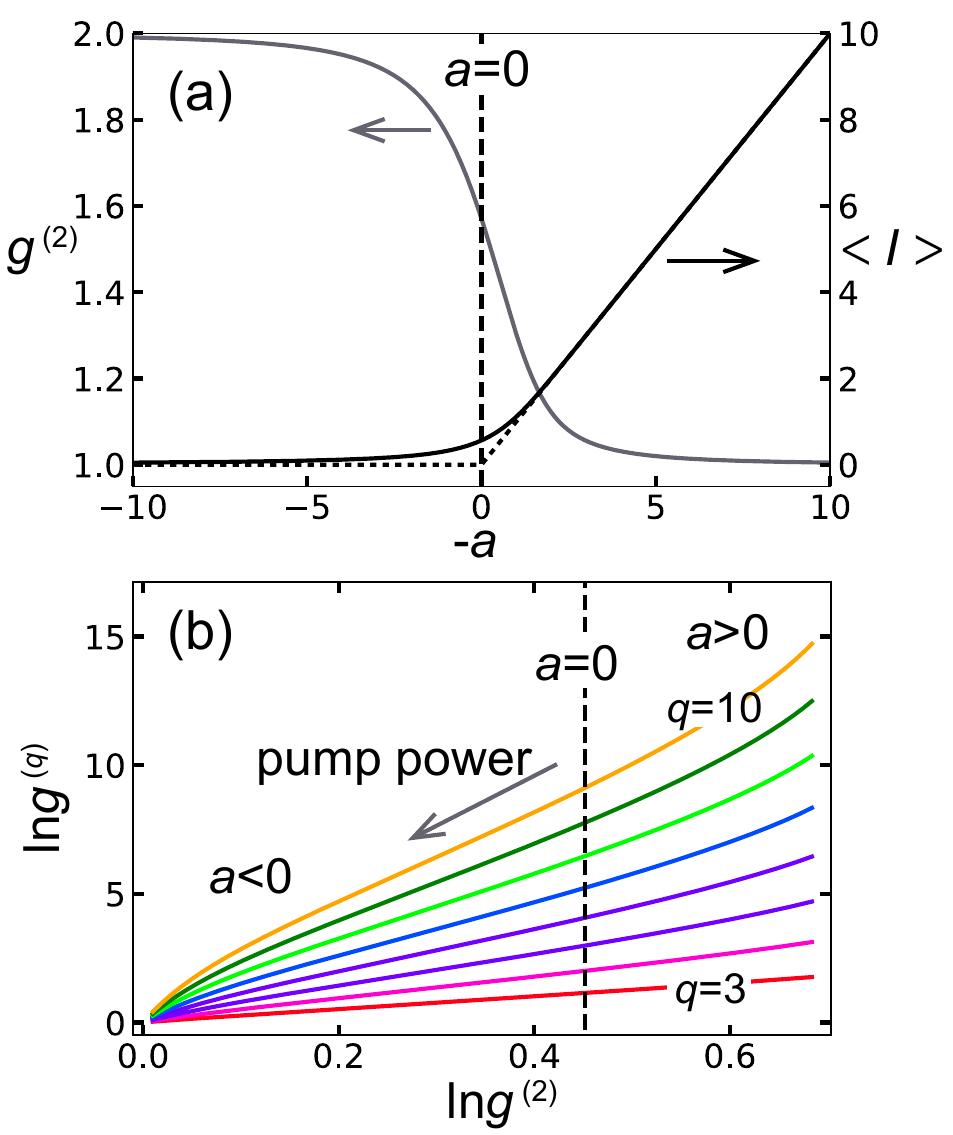}
\caption{(Color) (a) For $b=0.5$, photon number $\langle I\rangle=G^{(1)}$ (black line) and the second-order photon correlation $g^{(2)}$ (grey line) are calculated as a function of parameter $-a$ using the Ginzburg-Landau (GL) laser theory. The horizontal dashed line represents the lasing threshold when $a=0$. (b). $\ln g^{(2)}$ $(q\geq3)$ vs. $\ln g^{(3)}$ of the GL type transition are plotted for the range $-a/2b=-10$ to 10. The plots are based on Eq. (\ref{eq:GL_gq}).}
\label{fig:GL_theory}
\end{figure}

\subsection{A. GL theory for lasers and higher-order photon correlations}
For certain Langevin dynamics, let us assume that a steady-state distribution function for a complex electric field $\alpha=x+iy$ is given with a potential $F(\alpha)$:
\begin{eqnarray}
P(\alpha)=\frac{1}{Z}e^{-F(\alpha)},\label{Pfunction}
\end{eqnarray}
The normalization constant $Z$ is given by
\begin{eqnarray}
Z=\int d^2\alpha e^{-F(\alpha)},
\end{eqnarray}
where $d^2\alpha\equiv d{\rm Re}[\alpha]\cdot d{\rm Im}[\alpha]=dxdy$. Furthermore, for class-A lasers, we assume that the potential function $F(\alpha)$ can be given by the Ginzburg-Landau (GL) type around the lasing transition:
\begin{equation}
F(\alpha)=F_{\rm GL}(\alpha)\equiv a|\alpha|^2+b|\alpha|^4=aI+bI^2,
\label{GL_F}
\end{equation}
where $I\equiv|\alpha|^2$ is used and $b>0$. As is well known for the GL theory, $a>0$ represents the normal phase (or the trivial phase), where the potential minimum is located at $I=|\alpha|^2=0$. Meanwhile, $a<0$ is the ordered phase with a broken symmetry, where the potential has a Mexican-hat shape, and its minimum is at $I=|\alpha|^2=-a/2b$. We note that the GL type potential in Eq. (\ref{GL_F}) is homogeneous and does not have a ``kinetic term" such as $c(\nabla\alpha)^2$, which is because we are considering single-mode lasers. The ``kinetic term" may play an important role in inhomogeneous systems such as multi-mode lasers \cite{Haken1977} or coupled laser arrays \cite{Gartner2019}. The absence of the ``kinetic term" makes integration very easy.

From a distribution function $P(I)$ with $I=|\alpha|^2$, the photon counting statistics with a generalized quantum efficiency $\eta(\leq1)$ are calculated as \cite{Walls2007}
\begin{eqnarray}
p_n^\eta=\int_0^\infty dIP(I)\ \frac{(\eta I)^n}{n!}e^{-\eta I}.
\label{eq:pn_eta}
\end{eqnarray}
For the conversion of the coordinate, the polar coordinate $\alpha=re^{i\theta}$ and the relation $d^2\alpha=rdrd\theta$ with $I=r^2$ are used. Importantly, the generalized quantum efficiency $\eta(\leq1)$ includes various quantum efficiencies such as the quantum efficiency of detectors, optical losses, and detection time \cite{SCULLY1969}. Now, the $q$th order moment $G^{(q)}$ with a quantum efficiency $\eta$ is generally defined as
\begin{eqnarray}
G^{(q)}_\eta&\equiv&\langle n(n-1)\cdots(n-q+1)\rangle=\sum_{n=q}^\infty\frac{n!}{(n-q)!}p_n^\eta\nonumber\\
&=&\eta^q\int d^2\alpha(|\alpha|^2)^qP(\alpha)\nonumber\\
&=&\eta^q\int_0^\infty dI\ I^qP(I)=\eta^q\langle I^q\rangle.
\label{eq:Gq_general}
\end{eqnarray}
When the distribution is given by $P(\alpha)=e^{-F_{\rm GL}(\alpha)}/Z$, the $q$th order moment is calculated as
\begin{eqnarray}
G^{(q)}_\eta=\frac{\eta^q\Gamma(q+1)}{(\sqrt{2b})^q}\frac{D_{-q-1}(\frac{a}{\sqrt{2b}})}{D_{-1}(\frac{a}{\sqrt{2b}})}.
\label{eq:unnorm_Gq}
\end{eqnarray}
Here, we used the formula
\begin{eqnarray}
\int_0^\infty dx\ x^qe^{-\left(ax+bx^2\right)}=\frac{\Gamma(q+1)}{(\sqrt{2b})^{q+1}}e^{\left(\frac{a^2}{8b}\right)}D_{-q-1}({a}/{\sqrt{2b}}),\nonumber\\
\end{eqnarray}
where $\Gamma(x)$ is the gamma function, and $D_\nu(x)$ is the parabolic cylinder function. We note that the first-order moment $G^{(1)}_\eta$ is equivalent to the mean photon number $\langle I\rangle$ detected with a quantum efficiency $\eta$. Finally, the normalized (factorized) photon correlation function $g^{(q)}$ (factorial moment $F_q$ in Ref. \cite{Hwa1992}) is obtained as
\begin{eqnarray}
g^{(q)}\equiv\frac{G^{(q)}_\eta}{(G^{(1)}_\eta)^q}=\frac{q!D_{-q-1}(\frac{a}{\sqrt{2b}})[D_{-1}(\frac{a}{\sqrt{2b}})]^{q-1}}{[D_{-2}(\frac{a}{\sqrt{2b}})]^q}.\nonumber\\
\label{eq:GL_gq}
\end{eqnarray}
The important property of this normalized photon correlation function $g^{(q)}$ is its quantum efficiency independence \cite{Avenhaus2010}, which arises because the factor $\eta^q$ is cancelled out as a common factor  in the numerator and denominator. This is in contrast to the photon counting distribution $p_n^\eta$, which clearly depends on $\eta$. Since it is practically impossible to achieve $\eta=1$ in experiments, the normalized higher-order photon correlation function $g^{(q)}$ has a great advantage over the photon counting distribution $p_n$. Experimentally, $g^{(q)}$ could be directly measured with multiple Hanbury-Brown Twiss  interferometers or multichannel detectors \cite{Stevens2010,Elvira2011}. Furthermore, $g^{(q)}$ can be constructed from a measured imperfect photon counting distribution $p_n^{\eta}$ by using Eq. (\ref{eq:Gq_general}) \cite{Young1994,Wiersig2009,Schlottmann2018}. A linear photodetector would also be available for measuring the continuous distribution $P(I)$ \cite{Lien2001,Wang2015,Wang2017}, which gives $g^{(q)}$ [see Eq. (\ref{eq:Gq_general})]. In all cases, the requirement is that the time resolution of the detectors must be faster than the intensity coherence time.

First, in Fig. \ref{fig:GL_theory}(a), we plot the mean photon number $\langle I\rangle$ for $\eta=1$ and the normalized second-order photon correlation function $g^{(2)}$ as a function of $-a/(2b)$. Figure \ref{fig:GL_theory} indicates that the GL theory reproduces all the well-known behaviors of lasing transition, namely the buildup of the photon number and the transition of $g^{(2)}$ from 2 to 1 with an increase in pump power. Note that, at the threshold, 
\begin{equation}
g^{(2)}=\frac{2D_{-3}(0)D_1(0)}{[D_2(0)]^2}=\pi/2\ \ {\rm for}\ \ a=0 
\label{eq:threshold}
\end{equation}
holds.

Second, following \cite{Hwa1992,Hwa1993}, we plot $\ln g^{(q)}$ vs. $\ln g^{(2)}$ in Fig. \ref{fig:GL_theory}(b), where $q\geq3$ and $y\equiv a/\sqrt{2b}$ ranges from $y=-10$ to 10. Importantly, as long as the system is described by the GL theory, the curves given by $\ln g^{(q)}$ vs. $\ln g^{(2)}$ always hold independent of the value of $b$. In the original proposal by Hwa and Nazirov, focusing on the linearity of the curve around the threshold, they found a scaling law $g^{(q)}\propto (g^{(2)})^{\beta_q}$ with $\beta^q=(q-1)^{\nu}$ and $\nu\simeq1.3$ as a proof of the second-order quark gluon plasma phase transition \cite{Cao1996}. In this paper, instead of using the exponent $\nu$, we directly compare numerically simulated photon correlations with the analytically obtained curves $\ln g^{(q)}$ vs. $\ln g^{(2)}$. Theoretically, this method is very useful when we use the Langevin equations. Namely, while the distribution $P(\alpha)$ is difficult to simulate with the Langevin equations or the Fokker-Planck equation, stochastic simulation of $g^{(q)}$ is straightforward. Note that, in principle, we can reconstruct the shape of the photon statistics if we can obtain infinite orders of $g^{(q)}$, which is explained in Appendix B. 

\subsection{B. Langevin dynamics as a basis of the GL theory}
At the end of this section, we discuss the Langevin dynamics that gives the field distribution described by GL theory. Let us consider the complex field $\alpha=x+iy$ that obeys the Langevin equation
\begin{eqnarray}
\dot{\alpha}=\frac{1}{2}\mu\alpha-\frac{1}{2}\lambda|\alpha|^2\alpha+f_\alpha,
\label{eq:Hopf}
\end{eqnarray}
where a noise term $f_\alpha=f_x+if_y$ satisfies correlations
\begin{eqnarray}
\langle f_x(t)f_x(t')\rangle&=&Q\delta(t-t')\nonumber\\
\langle f_y(t)f_y(t')\rangle&=&Q\delta(t-t')\label{eq:SL_noise}\\
\langle f_x(t)f_y(t')\rangle&=&0\nonumber
\end{eqnarray}
and $\langle f_{x,y}(t)\rangle=0 $. Note that the deterministic part of Eq. (\ref{eq:Hopf}) is called the Stuart-Landau equation, which is known to exhibit the Hopf bifurcation when the control parameter $\mu$ becomes positive from negative \cite{Guckenheimer1984,Kuramoto2003}. Furthermore, the Stuart-Landau equation is interpreted as equilibrium motion in the GL potential. Therefore, the corresponding Fokker-Planck equation of motion satisfies the detailed-balance condition and can have a steady-state solution given by the GL potential as \cite{Risken1967,Haken2012,Risken1996,Lax1967,Louisell1973}
\begin{equation}
P(\alpha)=Z^{-1}\exp\left(-\frac{-\frac{1}{4}\mu|\alpha|^2+\frac{1}{8}\lambda|\alpha|^4}{\frac{1}{2}Q}\right).
\label{eq:P_GL}
\end{equation}
The derivation of Eq. (\ref{eq:P_GL}) from the Fokker-Planck equation of motion is detailed in Appendix A. Thus, for the Langevin dynamics represented by Eq. (\ref{eq:Hopf}), the parameters $a$ and $b$ in the GL potential are given  as $a=-\mu/(2Q)$ and $b=\lambda/(4Q)$, respectively. In terms of field intensity $I\equiv|\alpha|^2$, with noise term $f_I$ originating from $f_{x,y}$, Eq. (\ref{eq:Hopf}) is written as
\begin{eqnarray}
\dot{I}=\mu I-\lambda I^2+f_I,
\label{eq:trans_normal}
\end{eqnarray}
whose deterministic part is known as the normal form of the transcritical bifurcation \cite{Guckenheimer1984}. Actually, the transcritical bifurcation is shown as a dotted line in Fig. \ref{fig:GL_theory} (a). Since this paper focuses on the amplitude mode of lasers, in the rest part of the paper, we frequently use Eq. (\ref{eq:trans_normal}) in addition to Eq. (\ref{eq:Hopf}).
\begin{figure*}
\begin{center}
\includegraphics[width=1\textwidth]{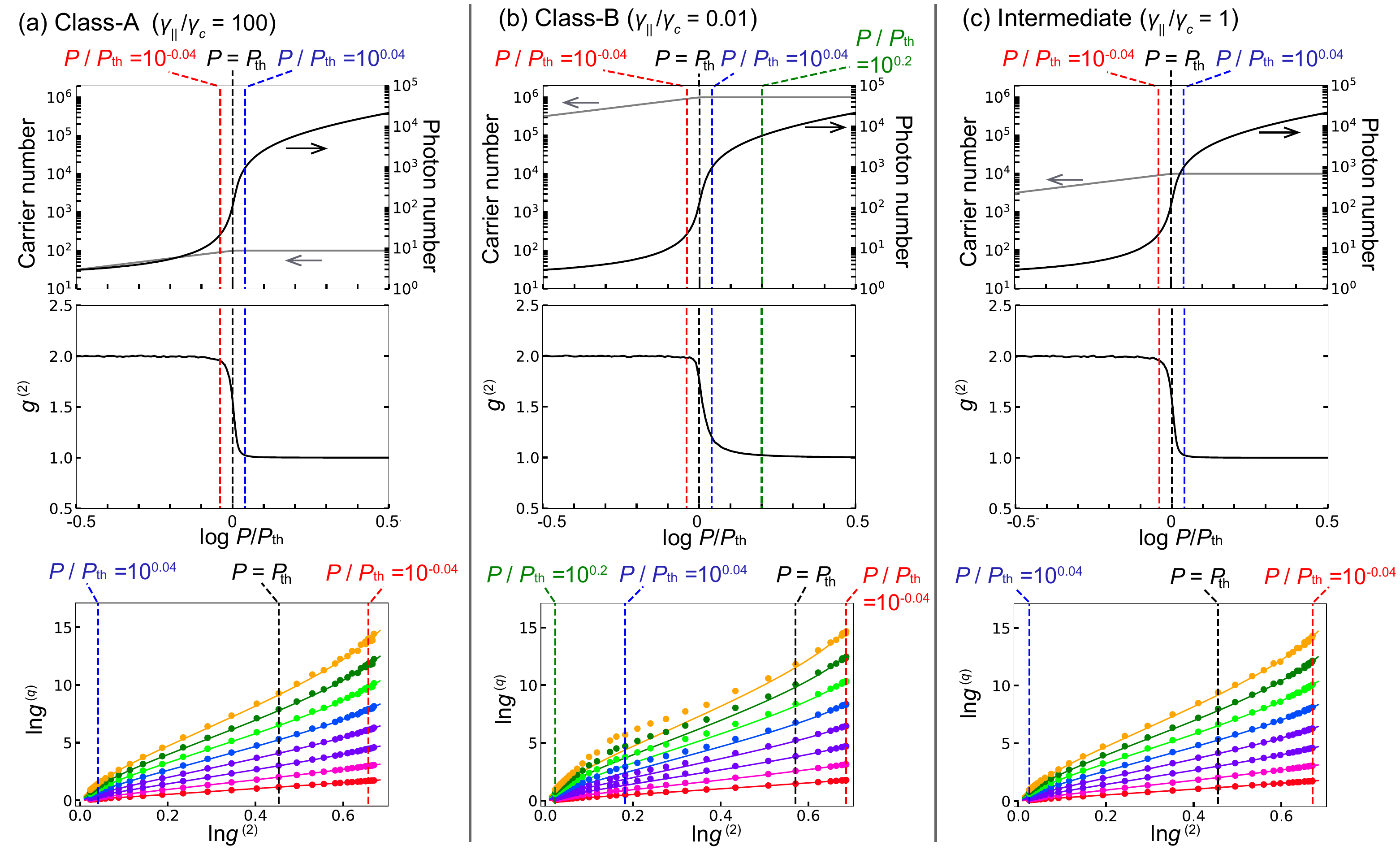}
\caption{(Color) Upper figures: photon number (right axis) and carrier number (left axis) plotted as a function of pump power. These plots are based on Eqs. (\ref{eq:REx})–(\ref{eq:REN}). Middle figures: second-order photon correlation function at zero time delay $g^{(2)}$ as a function of pump power. Bottom figures: plots of simulated $\ln g^{(q)}$ vs. $\ln g^{(2)}$, where $q$ ranges from 3 to 10 and the pump power increases from right to left. The solid curves show analytical $\ln g^{(q)}$ vs. $\ln g^{(2)}$ results based on the GL theory [Eq. (\ref{eq:GL_gq})]. Here, (a), (b), and (c) are respectively for the class-A ($\gamma_\|/\gamma_c=100$), class-B ($\gamma_\|/\gamma_c=0.01$), and intermediate ratio ($\gamma_\|/\gamma_c=1$) between the photon and carrier decay rates. For all the simulations, we used $\beta=10^{-4}$. The vertical dashed lines represent specific pump powers, $P/P_{\rm th}=10^{-0.04}$, 1, $10^{0.04}$, and $10^{0.2}$.}
\label{fig:classAB}
\end{center}
\end{figure*}

\section{4. Simulation}
\subsection{A. Stochastic rate equations}
Now, we describe a simulation technique for lasers with a wide range of parameters. The Statz-de Mars rate equations for continuous photon (field intensity) $I$ and carrier number (population inversion) $N$ are given by \cite{Bjork1991,Rice1994} 
\begin{eqnarray}
\dot{I}&=&F_I(I,N)=-\gamma_cI+\beta\gamma_\|NI\label{eq:REp_MF}\\
\dot{N}&=&F_N(I,N)=-\gamma_\|N-\beta\gamma_\|NI+P\label{eq:REn_MF}
\end{eqnarray}
with the pump power $P$.
Here, coefficient $\beta$ is referred to as the ``spontaneous emission coupling coefficient" representing the fraction of the spontaneous emission going into the cavity mode \cite{Bjork1991,Rice1994,Takemura2019,Takemura2021}. However, it should be noted that the spontaneous emission terms themselves are neglected in Eqs. (\ref{eq:REp_MF})–(\ref{eq:REn_MF}) because of $\beta\ll1$, and that, in this paper, $\beta$ should be interpreted simply as a photon-carrier coupling constant. The photon and population carrier decay rates are represented by and $\gamma_c$ and $\gamma_\|$, respectively. To obtain the above rate equations from the Maxwell-Bloch equations, the first requirement is a large dephasing rate $\gamma_\perp\gg\gamma_c,\gamma_\|$, which leads to the adiabatic elimination of the polarization degree of freedom (see the introduction). This condition is satisfied for almost all lasers. Second, we consider only low-$\beta$ lasers, $\beta\ll1$. When $\beta\ll1$, we can also neglect the carrier transparency number from the original rate equations proposed in Refs. \cite{Bjork1991,Rice1994} (see also Ref. \cite{Takemura2019}). This condition also holds for most lasers. 

Rate equations (\ref{eq:REp_MF}) and (\ref{eq:REn_MF}) indicate that lasing occurs at a threshold pump power $P=P_{\rm th}$, where $P_{\rm th}$ is given by
\begin{equation}
P_{\rm th}=\frac{\gamma_c}{\beta}.
\end{equation}
Below and above the lasing threshold, the steady-state solutions $I_0$ and $N_0$ are given by
\begin{equation}
I_0 =0\ \ {\rm and}\ \ N_0 =\frac{P}{\gamma_\|}\ \ {\rm for}\ \ P\leq P_{\rm th}.
\label{eq:below}
\end{equation}
and
\begin{equation}
I_0 =\frac{P}{\gamma_c}-\frac{1}{\beta}\ \ {\rm and}\ \ N_0 =\frac{\gamma_c}{\beta\gamma_\|}\ \ {\rm for}\ \ P>P_{\rm th},
\label{eq:above}
\end{equation}
respectively. Note that, as a function of the pump power $P$, $I_0$, and $N_0$ behave in a way similar to the dotted line in Fig. \ref{fig:GL_theory}(a).

Since we are interested in photon statistics, we add the Langevin noises $f_\alpha=f_x+if_y$ and $f_N$ to the field and carrier dynamics, respectively. In terms of the complex field $\alpha=x+iy$ for the rotating frame of the laser frequency, the rate equations (\ref{eq:REp_MF}) and (\ref{eq:REn_MF}) are written as \cite{Lariontsev2011} 
\begin{eqnarray}
\dot{x}&=&-\frac{1}{2}\gamma_cx+\frac{1}{2}\beta\gamma_\|Nx+f_x\label{eq:REx}\\
\dot{y}&=&-\frac{1}{2}\gamma_cy+\frac{1}{2}\beta\gamma_\|Ny+f_y\label{eq:REy}\\
\dot{N}&=&-\gamma_\|N-\beta\gamma_\|N(x^2+y^2)+P+f_N,\label{eq:REN}
\end{eqnarray}
where the field noise terms $f_x$ and $f_y$ satisfy the same correlations as in Eq. (\ref{eq:SL_noise}) with the field noise strength $Q_c$, while the carrier noise term $f_N$ follow the correlations
\begin{equation}
\langle f_N(t)f_N(t')\rangle=Q_N\delta(t-t')\ \ {\rm and}\ \ \langle f_N(t)\rangle=0.
\label{eq:carrier_noise}
\end{equation}
We call Eqs. (\ref{eq:REx})–(\ref{eq:REN}) stochastic rate equations. 

Before moving to the direct numerical simulation of the stochastic rate equations, we review the conventional laser theory for class-A lasers \cite{Risken1967,Haken2012,Risken1996,Lax1967,Louisell1973}, where the photon lifetime is much longer than the carrier (population inversion) lifetime $\gamma_c\ll\gamma_\|$. As mentioned in the introduction, we are able to eliminate the carrier degree of freedom from Eqs. (\ref{eq:REx})–(\ref{eq:REN}) by setting $\dot{N}=0$: 
\begin{eqnarray}
\bar{N}=\frac{{P}/\gamma_\|}{1+\beta (x^2+y^2)}\simeq\frac{P}{\gamma_\|}-\beta\frac{P}{\gamma_\|}(x^2+y^2),
\label{eq:adia_N}
\end{eqnarray}
where we also used the fact that $\beta\ll1$. This is the conventional adiabatic elimination procedure. Additionally, we neglected the noise term $f_N$ in adiabatic elimination, which is discussed again in Section 5E. Substituting Eq. (\ref{eq:adia_N}) into Eqs (\ref{eq:REx})–(\ref{eq:REN}), around the threshold, we obtain
\begin{eqnarray}
\dot{\alpha}&=&-\frac{1}{2}\gamma_c\alpha+\frac{1}{2}\beta P\alpha+\frac{\gamma_c}{2}\beta\frac{P}{P_{\rm th}}|\alpha|^2\alpha+f_\alpha\nonumber\\
&\simeq&\frac{\gamma_c}{2}\epsilon\alpha-\beta\frac{\gamma_c}{2}|\alpha|^2\alpha+f_\alpha
\label{eq:van_der_Pol}
\end{eqnarray}
with a pump parameter defined as
\begin{equation}
\epsilon\equiv\frac{P-P_{\rm th}}{P_{\rm th}}.
\label{eq:pump_epsilon}
\end{equation}  
where we used approximation $P/P_{\rm th}\simeq1$, which holds around the lasing threshold $P\simeq P_{\rm th}$. Equation (\ref{eq:van_der_Pol}) is clearly the same as Eq.  (\ref{eq:Hopf}). Therefore, when $\gamma_c\ll\gamma_\|$ (class-A lasers), the steady-state field distribution of lasers is given by the GL potential Eq. (\ref{GL_F}). Therefore, the light output intensity and the second-order correlation $g^{(2)}$ of class-A lasers are expected to be the same as those obtained with the GL theory shown in Fig. \ref{fig:GL_theory}(a). This is the outline of the laser-phase transition analogy described in Refs. \cite{Haken2012,Scully1999}. Here, the surprise is that non-equilibrium rate equations are transformed to the equilibrium dynamics. Namely, the original stochastic rate equations (\ref{eq:REx})–(\ref{eq:REN}) clearly have pump and dissipation, and thus the corresponding Fokker-Planck equation obviously violates the detailed balance condition. Meanwhile, the field equation of motion after adiabatic elimination, Eq. (\ref{eq:van_der_Pol}), represents equilibrium motion in the GL potential, and as discussed in \cite{Risken1996} (see also Appendix A), the corresponding Fokker-Planck equation satisfies the detailed-balance condition in the rotating frame of the laser frequency. Therefore, in rewriting the non-equilibrium model to the equilibrium one, the adiabatic elimination plays a key role.

The GL theory is well established for low-$\beta$ class-A lasers. However, whether there is an actual limit of the ratio $\gamma_\|/\gamma_c$ for the applicability of the GL theory is not clear. In the next subsection, we numerically simulate the stochastic rate equations (\ref{eq:REx})–(\ref{eq:REN}) for various ratios $\gamma_\|/\gamma_c$, and check whether or not their photon statistics are described by the GL theory using higher-order photon correlations. 

\subsection{B. Results}
Here, we present numerical simulations of the stochastic rate equation (\ref{eq:REx})–(\ref{eq:REN}) and compare simulated photon correlation functions with the predictions of the GL theory [see Eq. (\ref{eq:GL_gq})]. For numerical simulations, we used the conventional Euler-Maruyama method. The higher-order photon correlations $g^{(q)}$ are numerically calculated as classical statistical averages $g^{(q)}=\langle I^q\rangle/\langle I\rangle^q$ [see Eq. (\ref{eq:unnorm_Gq})], where $I=x^2+y^2$. In all the simulations, for simplicity, we used the same noise strengths for the field and carrier:
\begin{equation}
Q_c=Q_N=\gamma_c.
\end{equation}
Furthermore, for all the simulations, we used $\beta=10^{-4}$, which satisfies $\beta\ll1$. Fig. \ref{fig:classAB}(a)–(c) show the results of the stochastic simulations for class-A ($\gamma_\|/\gamma_c=100$), B ($\gamma_\|/\gamma_c=0.01$) and intermediate ($\gamma_\|/\gamma_c=1$) parameters, , respectively. These are  the central results of this paper. The top row shows the photon and carrier numbers as a function of pump power.  The pump-input and light-output curves in Fig. \ref{fig:classAB}(a)–(c) (see the top row) exhibit sharp kinks associated with a very small $\beta(=10^{-4})$. The middle row in Fig. \ref{fig:classAB} plots the second-order photon correlation at a zero time delay [$g^{(2)}$] as a function of pump power. In the bottom row in Fig. \ref{fig:classAB}, we plot $\ln g^{(q)}$ vs. $\ln g^{(2)}$ around the lasing transition, where the colored solid lines are the analytical results for the GL theory [see Eq. (\ref{eq:GL_gq}) and Fig. \ref{fig:GL_theory}(b)].

First, we examine a laser with $\beta=10^{-4}$ and $\gamma_\|/\gamma_c=100$ (a class-A laser) [see Fig. \ref{fig:classAB}(a)]. The second-order photon correlation $g^{(2)}$ shows a sharp transition from $g^{(2)}=2$ to 1 at the lasing threshold ($P=P_{\rm th}$). In the bottom row in Fig. \ref{fig:classAB}(a), all the simulated results (filled colored circles) fall on the solid colored curves, which indicates that the photon statistics of the class-A laser are well described by the GL theory. In fact, this is what we expect from the argument from Eq. (\ref{eq:adia_N}) and (\ref{eq:van_der_Pol}). 

Second, we investigate a laser with $\beta=10^{-4}$ and $\gamma_\|/\gamma_c=0.01$ (a class-B laser) [see Fig. \ref{fig:classAB}(b)]. Interestingly, $g^{(2)}$ in Fig. \ref{fig:classAB}(b) behaves very differently from that in Fig. \ref{fig:classAB}(a). Namely, the super Poissonian photon bunching [$g^{(2)}>1$] remains above the lasing threshold. This long-tailed bunching behavior of $g^{(2)}$ at high pump power is characteristic behavior of class-B lasers and has been experimentally reported in Refs. \cite{Druten2000,Takemura2012,Wang2017,Takemura2019}. However, the information on $g^{(2)}$ is insufficient to characterize the full photon statistics. Therefore, we use the information on higher-order photon correlation functions $\ln g^{(q>2)}$ vs. $\ln g^{(2)}$, which clearly indicates that the filled colored circles deviate from the prediction s of the GL theory (solid colored curves). Therefore, the lasing transition with the class-B parameters cannot be described by the GL theory. 

Finally, we discuss the intermediate parameter with $\beta=10^{-4}$ and $\gamma_\|/\gamma_c=1$ [See Fig. \ref{fig:classAB}(c)]. Surprisingly, even though the ratio $\gamma_\|/\gamma_c=1$ does not satisfy the conventional adiabatic elimination condition $\gamma_\|/\gamma_c\gg1$ at all, all the simulated results are almost identical to those in Fig. \ref{fig:classAB}(a). Namely, the second-order photon correlation $g^{(2)}$ exhibits a sharp drop from $g^{(2)}=2$ to 1 at the lasing threshold, and $\ln g^{(q)}$ vs. $\ln g^{(2)}$ are plotted on the solid curves, which are the predictions of the GL theory. This result indicates that the applicability of the GL theory of lasers is broader than conventionally imagined.

\section{5. Analysis}
In this section, we demonstrate that, under a certain condition, the deterministic photon and carrier rate equations (\ref{eq:REp_MF}) and (\ref{eq:REn_MF}) can be reduced to a single equation of motion:
\begin{equation}
\dot{I}=\gamma_c\epsilon I-\beta\gamma_c I^2,
\label{eq:transcritical}
\end{equation}
which has already been introduced in Eq. (\ref{eq:trans_normal}). For this purpose, we extend the conventional adiabatic elimination method (see Section 4A) by using the center manifold reduction theory \cite{Carr2012,Guckenheimer1984,Wunderlin1981}.
\begin{figure}
\includegraphics[width=0.35\textwidth]{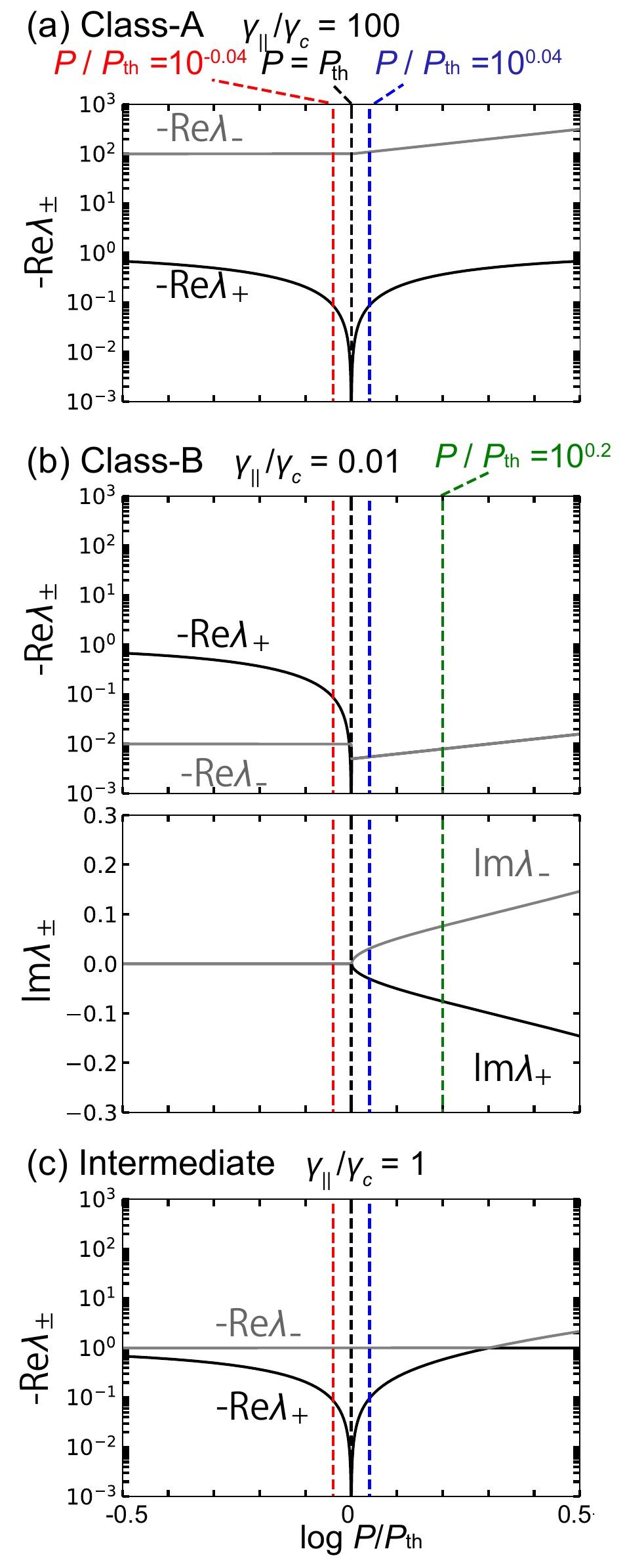}
\caption{(Color) Real ${\rm Re}\lambda_\pm$ and imaginary part ${\rm Im}\lambda_\pm$ of the eigenvalues of the Jacobian ${\bm L}$ [Eq. (\ref{eq:matrix_d})] as a function of pump power. (a), (b), and (c) are for class-A ($\gamma_\|/\gamma_c=100$), class-B ($\gamma_\|/\gamma_c=0.01$), and intermediate parameters ($\gamma_\|/\gamma_c=1$), respectively. For all the simulations, we used $\beta=10^{-4}$. Note that the imaginary part ${\rm Im}\lambda_\pm$ is always zero except for  (b).}
\label{fig:eigenvalues}
\end{figure}

\subsection{A. Linear stability analysis}
First, we perform a linear stability analysis. The small fluctuations $\delta I$ and $\delta N$ around the steady states defined as $I=\bar{I} +\delta I$ and $N=\bar{N} +\delta N$, respectively, follow the equation of motion
\begin{eqnarray}
&&
\frac{d}{dt}
\left( \begin{array}{c}
\delta I\\
\delta N\\
\end{array} \right)
={\bm L}
\left( \begin{array}{c}
\delta I\\
\delta N\\
\end{array} \right)
+
{\bm G},\label{eq:matrix}
\end{eqnarray}
and the matrix for the linear part ${\bm L}$ (Jacobian) is
\begin{eqnarray}
{\bm L}=
\left( \begin{array}{cc}
\frac{\partial F_I}{\partial I} & \frac{\partial F_I}{\partial N}\\[5pt]
\frac{\partial F_N}{\partial I} & \frac{\partial F_N}{\partial N}\\
\end{array} \right)
=
\left( \begin{array}{cc}
-\gamma_c+\beta\gamma_\|\bar{N}  & \beta\gamma_\|\bar{I} \\[5pt]
-\beta\gamma_\|\bar{N}  & -\gamma_\|-\beta\gamma_\|\bar{I} \\
\end{array} \right)\nonumber\\\label{eq:matrix_d}
\end{eqnarray}
and the nonlinear part ${\bm G}$ is given by
\begin{eqnarray}
&&
{\bm G}
=
\left( \begin{array}{c}
\beta\delta I\delta N-\gamma_c\bar{I} +\beta\gamma_\|\bar{N} \bar{I} \\[5pt]
-\beta\delta I\delta N-\gamma_\|\bar{N} -\beta\gamma_\|\bar{N} \bar{I} +P\\
\end{array} \right).
\end{eqnarray}
We calculate the eigenvalues of the Jacobian ${\bm L}$. For this purpose, Eq. (\ref{eq:matrix}) is further simplified depending on whether pump power is below or above the lasing threshold.

(i)  Below the lasing threshold $\epsilon\leq0$ ($P\leq P_{\rm th}$): substituting $\bar{I} =0$ and $\bar{N} ={P}/{\gamma_\|}$ [see Eq. (\ref{eq:below})] into Eq. (\ref{eq:matrix}), we obtain the eigenvalues of the Jacobian ${\bm L}$ as
\begin{equation}
\lambda_+=\gamma_c\epsilon\ \ {\rm and}\ \ \lambda_-=-\gamma_\|\ \ {\rm for}\ \ \epsilon\leq 0.\label{eq:eigen_below}
\end{equation}
Note that the eigenvalues are always real below the lasing threshold for any ratio $\gamma_\|/\gamma_c$.

(ii)  Above the lasing threshold $\epsilon>0$ ($P>P_{\rm th}$): substituting $\bar{I} =P/\gamma_c-1/\beta=\epsilon/\beta$ and $\bar{N} ={\gamma_c}/{\beta\gamma_\|}$ [see Eq. (\ref{eq:above})] into Eq. (\ref{eq:matrix}), we obtain the eigenvalues of the Jacobian ${\bm L}$ as
\begin{eqnarray}
\lambda_\pm&=&\frac{1}{2}\left[-\gamma_\|(\epsilon+1)\pm\sqrt{\gamma_\|^2(\epsilon+1)^2-4\gamma_c\gamma_\|\epsilon}\right]\nonumber\\
&&\ \ \ \ \ \ \ \ \ \ \ \ \ \ \ \ \ \ \ \ \ \ \ \ \ \ \ \ \ \ \ \ \ \ \ \ \ \ \ {\rm for}\ \ \epsilon>0.\label{eq:eigen_above}
\end{eqnarray}
Importantly, above the lasing threshold, the eigenvalues $\lambda_\pm$ can be complex when $\gamma_\|/\gamma_c<1$. If the inside of the square root of Eq. (\ref{eq:eigen_above}) is negative, the eigenvalues are written as
\begin{equation}
\lambda_\pm=-\gamma_{\rm ro}\mp i\omega_{\rm ro}\ \ {\rm for}\ \ \gamma_\|(\epsilon+1)^2<4\gamma_c\epsilon,
\label{eq:eigen_ro}
\end{equation}
where $\gamma_{\rm ro}=\gamma_\|\epsilon$ and $\omega_{\rm ro}=\sqrt{4\gamma_c\gamma_\|\epsilon-\gamma_\|^2(\epsilon+1)^2}$ are interpreted as the damping rate and the oscillation frequency of the photon-carrier relaxation oscillation, respectively \cite{Takemura2012,Wang2015}.

In Fig. \ref{fig:eigenvalues}, we plot the eigenvalues $\lambda_\pm$ as a function of pump power for the three  different parameters: class-A $\gamma_\|/\gamma_c=0.01$ (a), class-B $\gamma_\|/\gamma_c=0.01$ (b), and the intermediate ratio $\gamma_\|/\gamma_c=1$ (c). Note that Eqs (\ref{eq:eigen_below}) and (\ref{eq:eigen_above}) indicate that $\lambda_+$ reaches zero at the lasing threshold $P=P_{\rm th}$ for any ratio $\gamma_\|/\gamma_c$. As we can expect from Eq. (\ref{eq:eigen_above}), above the lasing threshold of the class-B laser, the real parts of the two eigenvalues degenerate and the imaginary parts appear [see Fig. \ref{fig:eigenvalues}(b)], while this does not occur for the class-A [see Fig. \ref{fig:eigenvalues}(a)] and the intermediate parameters [see Fig. \ref{fig:eigenvalues}(c)]. 

From Fig. \ref{fig:eigenvalues}, we can naively guess that, for the class-A (a) and intermediate parameters (c), adiabatic elimination may be applicable in the broad region around where $-{\rm Re}\lambda_+\ll -{\rm Re}\lambda_-$ holds. Meanwhile, for the class-B parameter (b), adiabatic elimination may be applicable only in the extremely narrow region where $-{\rm Re}\lambda_+\ll -{\rm Re}\lambda_-$ holds. In the next subsection, we attempt to verify these naive expectations using the center manifold reduction theory. We also show that the imaginary parts of the eigenvalues associated with the photon-carrier relaxation oscillation play a central role in the breakdown of the GL theory.

\subsection{B. Center manifold reduction}
Our objective is to find, if it exists, a manifold (curve) that works as an attractor for the motion. The two-dimensional motion of $\delta I$ and $\delta N$ may be reduced to a one-dimensional motion on the attractor curve, which is the central idea of center manifold reduction \cite{Carr2012,Oppo1986}. Importantly, here, we treat the pump parameter $\epsilon$ also as a variable, and this is called the suspension trick \cite{Carr2012,Wunderlin1981}. In the same way as we did for the photon and carrier numbers, we separate $\epsilon$ into the ``mean value" and fluctuation as $\epsilon=\bar{\epsilon}+\delta \epsilon$. In lasers, the variable $\delta \epsilon$ is more than a mathematical trick because $\delta \epsilon$ may represent the pump power fluctuation. First, we reinterpret Eqs. (\ref{eq:REp_MF}) and (\ref{eq:REn_MF}) as the equations of motion for $I$, $N$, and $\epsilon$:
\begin{eqnarray}
\dot{I}&=&F_I(I,N,\epsilon)=-\gamma_cI+\beta\gamma_\|NI\label{eq:REp_sus}\\
\dot{N}&=&F_N(I,N,\epsilon)=-\gamma_\|N-\beta\gamma_\|NI+\frac{\gamma_c}{\beta}(\epsilon+1)\label{eq:REn_sus}\\
\dot{\epsilon}&=&F_\epsilon(I,N,\epsilon)=0\label{eq:REe_sus}.
\end{eqnarray}
Now, importantly, we consider small fluctuations as $\delta I$, $\delta N$, and $\delta \epsilon$ around the lasing threshold. Therefore, the mean values of $\bar{I}$, $\bar{N}$, and $\bar{\epsilon}$ are given at the lasing threshold $P=P_{\rm th}$ respectively as 
\begin{equation}
\bar{I} =0,\ \ {\rm }\ \ \bar{N} =\frac{P_{\rm th}}{\gamma_\|}=\frac{\gamma_c}{\beta\gamma_\|},\ \ {\rm and}\ \  \bar{\epsilon}=0.
\end{equation} 
Now, the equations of motion for the fluctuations are given by
\begin{eqnarray}
\frac{d}{dt}
\left( \begin{array}{c}
\delta I\\
\delta N\\
\delta\epsilon\\
\end{array} \right)
&=&
\left( \begin{array}{ccc}
0 & 0 & 0\\
-\gamma_c & -\gamma_\| &\gamma_c/\beta\\
0 & 0 &0\\
\end{array} \right)
\left( \begin{array}{c}
\delta I\\
\delta N\\
\delta\epsilon\\
\end{array} \right)\nonumber\\
&&\ \ \ \ \ \ \ \ \ \ +
\left( \begin{array}{c}
\beta\gamma_\|\delta I\delta N\\
-\beta\gamma_\|\delta I\delta N\\
0\\
\end{array} \right).
\label{eq:suspended}
\end{eqnarray}
The matrix of the linear part has eigenvalues $0$, $0$, and $-\gamma_\|$. Since these eigenvalues are zero and negative real values , the center manifold theorem guarantees the existence of a center manifold \cite{Carr2012}. By defining a new variable $\delta v$ as
\begin{equation}
\delta v=\frac{\gamma_c}{\gamma_\|}\delta I+\delta N-\frac{\gamma_c}{\beta\gamma_\|}\delta\epsilon,
\end{equation}
the equation of motion Eq. (\ref{eq:suspended}) is transformed to
\begin{eqnarray}
\delta\dot{I}&=&\gamma_c\delta\epsilon\delta I-\beta\gamma_c\delta I^2+\beta\gamma_\|\delta I\delta v\label{eq:deltaI_v}\\
\delta\dot{v}&=&-\gamma_\|v-\beta(\gamma_\|-\gamma_c)\delta I\delta v\nonumber\\
&&+\beta(\gamma_\|-\gamma_c)\frac{\gamma_c}{\gamma_\|}\delta I^2-(\gamma_\|-\gamma_c)\frac{\gamma_c}{\gamma_\|}\delta\epsilon\delta I\\
\delta\dot{\epsilon}&=&0.
\end{eqnarray}
Since $\delta I$ and $\delta \epsilon$ have finite decay rates, while $\delta v$ has a zero decay rate, $\delta I$ and $\delta\epsilon$ are referred to as unstable modes that ``enslave" the stable mode $\delta v$ \cite{Haken2012,Haken2012a}. According to Haken's ``slaving principle", the unstable modes $\delta I$ and $\delta\epsilon$ are also called ``order parameters" that govern the slow dynamics of the system. Here, the center manifold $\delta v=h(\delta I,\delta\epsilon)$ is obtained as
\begin{eqnarray}
\delta v&=&h(\delta I,\delta\epsilon)=\beta(\gamma_\|-\gamma_c)\frac{\gamma_c}{\gamma_\|^2}\delta I^2\nonumber\\
&&-(\gamma_\|-\gamma_c)\frac{\gamma_c}{\gamma_\|^2}\delta\epsilon\delta I+\mathcal{O}(C(\delta I,\delta\epsilon)),
\label{eq:CM_vbasis}
\end{eqnarray}
where $C(x,y)$ represents a homogeneous cubic in terms of $x$ and $y$. In the basis $\delta N$, the center manifold $\delta N=\tilde{h}(\delta I,\delta\epsilon)$ is written as
\begin{eqnarray}
\delta N&=&\tilde{h}(\delta I,\delta\epsilon)\nonumber\\
&=&-\frac{\gamma_c}{\gamma_\|}\delta I+\frac{\gamma_c}{\beta\gamma_\|}\delta\epsilon+\beta(\gamma_\|-\gamma_c)\frac{\gamma_c}{\gamma_\|^2}\delta I^2\nonumber\\
&&-(\gamma_\|-\gamma_c)\frac{\gamma_c}{\gamma_\|^2}\delta\epsilon\delta I+\mathcal{O}(C(\delta I,\delta\epsilon)).
\label{eq:CM_Nbasis}
\end{eqnarray}
Finally, substituting $\delta v=h(\delta I,\delta\epsilon)$ [Eq. (\ref{eq:CM_vbasis})] in the equation of motion Eq. (\ref{eq:deltaI_v}), we obtain the reduced equation of motion solely of photon fluctuation:
\begin{equation}
\delta \dot{I}=\gamma_c\delta\epsilon\delta I-\beta\gamma_c\delta I^2+\mathcal{O}(C(\delta I,\delta\epsilon)),\label{eq:reduced}
\end{equation}
where $\delta I$ and $\delta\epsilon$ can be replaced respectively with $I$ and $\epsilon$ because $\bar{I}=0$ and $\bar{\epsilon}=0$. The above equation of motion is nothing else but Eq. (\ref{eq:transcritical}) that exhibits transcritical bifurcation. Interestingly, the reduction of dynamics to the slow equation of motion (\ref{eq:transcritical}) is always possible around the neighborhood of the lasing threshold for any ratio $\gamma_\|/\gamma_c$. However, the above analysis does not provide the actual range of the neighborhood, which strongly depends on the ratio $\gamma_\|/\gamma_c$. Thus, in the next subsection, we discuss the applicability of the center manifold for the three different ratios $\gamma_\|/\gamma_c$ using the phase portraits.
\begin{figure*}
\begin{center}
\includegraphics[width=1\textwidth]{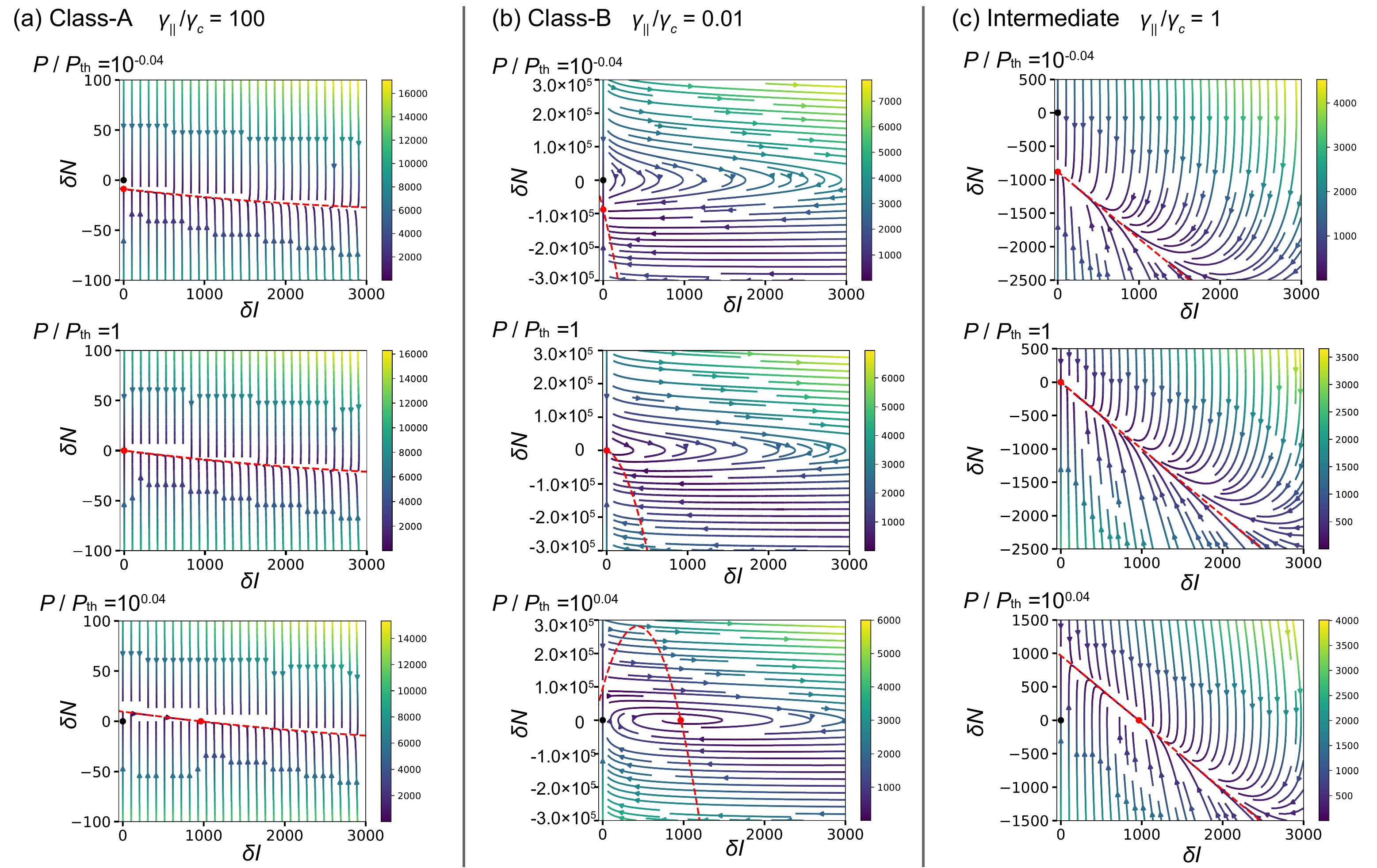}
\caption{(Color) Phase portraits are plotted based on Eq. (\ref{eq:portrait}), where (a), (b), and (c) are for class-A ($\gamma_\|/\gamma_c=100$), class-B ($\gamma_\|/\gamma_c=0.01$), and intermediate parameters ($\gamma_\|/\gamma_c=1$), respectively. For all simulations, $\beta=10^{-4}$ was used. Top, middle, and bottom figures represent results obtained below ($P/P_{\rm th}=10^{-0.04}$), at ($P/P_{\rm th}=1$), and above the lasing threshold ($P/P_{\rm th}=10^{0.04}$). The colors of the arrows represent the speeds of the flows, $(\dot{I}^2+\dot{N}^2)^{1/2}$. The red dashed curves represent the center manifold given by Eq. (\ref{eq:CM_Nbasis}). The red filled circles represent fixed points of the flows.}
\label{fig:flow}
\end{center}
\end{figure*} 

\subsection{C. Phase portraits}
Figure \ref{fig:flow} shows the phase portraits of the photon and carrier fluctuations represented as a vector ${\bf v}=(\delta \dot{I}, \delta \dot{N})$ defined as
\begin{eqnarray}
&&{\bf v}=
\left( \begin{array}{c}
\delta \dot{I}\\
\delta \dot{N}\\
\end{array} \right)
=
\left( \begin{array}{c}
\beta\delta I\delta N\\
-\gamma_c\delta I-\gamma_\|\delta N+(\gamma_c/\beta)\delta\epsilon-\beta\delta I\delta N\\
\end{array} \right),\nonumber
\label{eq:portrait}\\
\end{eqnarray}
which is obtained by fixing $\delta\epsilon(=\epsilon)$. The phase portraits are for (a) class-A ($\gamma_\|/\gamma_c=100$), (b) class-B ($\gamma_\|/\gamma_c=0.01$), and (c) the intermediate ($\gamma_\|/\gamma_c=1$) parameters. The top, middle, and bottom phase portraits show results obtained below ($P/P_{\rm th}=10^{-0.04}$), at ($P/P_{\rm th}=1$), and above the lasing threshold ($P/P_{\rm th}=10^{0.04}$), respectively. Note that we used the relation $\delta\epsilon=\epsilon=P/P_{\rm th}-1$ [Eq. (\ref{eq:pump_epsilon}]. The origins of the coordinate are plotted by filled blue circles, while the red filled circles represent the fixed points of the flows, where ${\bf v}=(\delta \dot{I}, \delta \dot{N})=(0,0)$ holds. The fixed point is given by $(0,\epsilon\gamma_c/(\beta\gamma_\|))$ below the lasing threshold ($\epsilon\leq0$), while it is given by $(\epsilon/\beta,0)$ above the lasing threshold ($\epsilon>0$). As we expected, Fig. \ref{fig:flow} indicates that this fixed point attracts the flow. The red dashed curves represent the center manifolds given by Eq (\ref{eq:CM_Nbasis}).

For the class-A and intermediate parameters [see Fig. \ref{fig:flow}(a) and (b)], the results indicate that the center manifolds (see the red dashed curves) work as attractors for the flows ${\bf v}$ for all three pump powers. Namely, the fluctuations $\delta I$ and $\delta N$ are rapidly attracted to the red dashed curves, and thus the slow dynamics can be described by the center manifolds. Furthermore, these results indicate that the fluctuations $\delta I$, $\delta N$, and $\delta \epsilon$ have a large neighborhood where the center manifolds can be applied. Even though, from the two-dimensional flows in Fig. \ref{fig:flow}, we cannot find the neighborhood of $\delta \epsilon$, we can find that the center manifold reduction is valid at least in the range from $1+\delta\epsilon=P/P_{\rm th}=10^{-0.1}$ to $10^{0.1}$. 

\begin{figure}
\includegraphics[width=0.35\textwidth]{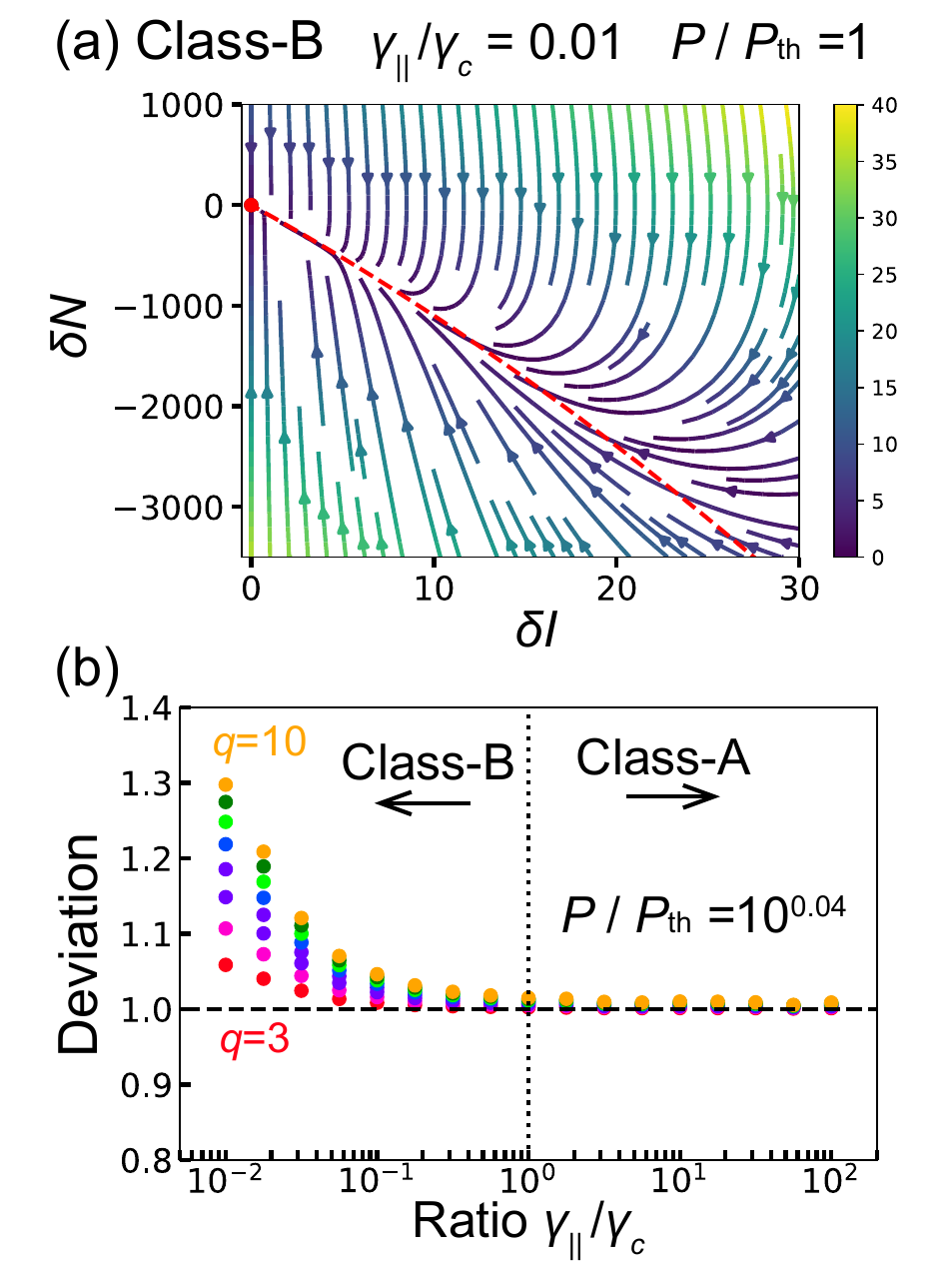}
\caption{(Color) (a) Zoomed-in phase portrait of the middle row in Fig. \ref{fig:flow}(b), which is for the class-B laser ($\gamma_\|/\gamma_c=0.01$) at the lasing threshold ($P/R_{\rm th}=1$). (b) The difference between the simulated photon correlations and the GL theory as a function of the ratio $\gamma_\|/\gamma_c$. For a simulated $\ln g^{(2)}_{\rm sim}$ at $P/P_{\rm th}=10^{0.04}$, we calculated deviation as $\ln g^{(q)}_{\rm sim}/\ln g^{(q)}_{\rm GL}$, where $\ln g^{(q)}_{\rm sim}$ is a simulated result, while $\ln g^{(q)}_{\rm GL}$ is an analytical result based on the GL theory [Eq. (\ref{eq:GL_gq})].}
\label{fig:suppl}
\end{figure}

On the other hand, for the class-B parameter, the phase portraits are strikingly different from those of the class-A and intermediate parameters. From Fig. \ref{fig:flow}(b), it is clear that the center manifolds (see the red dashed curve) do not work as attractors for the flow. In particular, in the bottom portrait in Fig. \ref{fig:flow}(b), due to the spiral focus, the center manifold reduction fails. This spiral focus of the flow originates from the imaginary parts of the eigenvalues ${\rm Im}\lambda_{\pm}$ [see Eq. (\ref{eq:eigen_ro})], which represents the photon-carrier relaxation oscillation in class-B lasers. In this sense, it is when there is photon-carrier relaxation oscillation that the center manifold reduction fails, because we fail to separate the time scales of field and carrier dynamics due to the mixing of the two dynamics. At the lasing threshold, special attention is required about the validity of the center manifold reduction. At a glance, in the middle graph  in Fig. \ref{fig:flow}(b), the center manifold (red dashed lines) does not work as an attractor. Additionally, it indicates that, at the lasing threshold $P=P_{\rm th}$, $g^{(2)}$ is larger than $\pi/2$, which also indicates the failure of center manifold reduction. However, as we discussed in the previous subsection, at the lasing threshold ($\epsilon=0$), center manifold reduction must always be valid regardless of the ratio $\gamma_\|/\gamma_c$. To solve this paradox, in Fig. \ref{fig:suppl}(a), we show a zoomed-in phase portrait for the class-B laser ($\gamma_\|/\gamma_c=0.01$) at the lasing threshold ($P=P_{\rm th}$), which clearly indicates that the center manifold works as an attractor only in the small region where $\delta I\lesssim10$. Therefore, in the stochastic system, photon noise may easily take the system out of the region where the center manifold reduction is valid.

\subsection{D. Transition from Class-B to Class-A laser}
Now, we address the important question: Where is the true boundary between the GL and non-GL photon statistics? Our important finding is that the unconventional photon statistics are associated with the nonzero imaginary part of the eigenvalue $\lambda_{\pm}$ (or the photon-carrier relaxation oscillation), where the center manifold reduction fails. As is evident from the inside of the square root of Eq. (\ref{eq:eigen_above}), the eigenvalue $\lambda_{\pm}$ has a nonzero imaginary part when $\gamma_\|/\gamma_c>1$; otherwise, $\lambda_{\pm}$ is a real value. Therefore, for low-$\beta$ lasers, we can say that the boundary between the GL and non-GL photon statistics is $\gamma_\|/\gamma_c=1$. Namely, the photon statistical properties of low-$\beta$ lasers with the ratio $\gamma_\|/\gamma_c\geq1$ can be described by the GL theory. Meanwhile, low-$\beta$ lasers with the ratio $\gamma_\|/\gamma_c<1$ exhibit non-GL photon statistics. 

To supplement this idea, in Fig. \ref{fig:suppl}(b), we show how photon statistics deviate from the prediction of the GL theory depending on the ratio $\gamma_\|/\gamma_c$. Figure \ref{fig:suppl}(b) shows how the plot $\ln g^{(q)}$ vs. $\ln g^{(2)}$ deviates from the prediction of the GL theory at a pump power of $P/P_{\rm th}=10^{0.04}$ for various ratios $\gamma_\|/\gamma_c$. The plots in Fig. \ref{fig:suppl}(b) were obtained as follows. For a simulated value of $\ln g^{(2)}_{\rm sim}$ for $P/P_{\rm th}=10^{0.04}$, the GL theory gives $\ln g^{(q)}_{\rm GL}$ with Eq. (\ref{eq:GL_gq}). Thus, we divide the simulated $\ln g^{(q)}_{\rm sim}$ by the GL prediction $\ln g^{(q)}_{\rm GL}$, which represents the difference between the simulation and the GL theory. Except for the ratio $\gamma_\|/\gamma_c$, the other parameters are the same as in Fig. \ref{fig:classAB}. We chose $P/P_{\rm th}=10^{0.04}$ as a characteristic pump power because it is close to the threshold but high enough for photon-carrier relaxation oscillation to appear. Figure \ref{fig:suppl}(b) clearly shows that the deviation from the GL theory starts to appear when $\gamma_\|/\gamma_c$ becomes smaller than unity. Therefore, the ratio $\gamma_\|/\gamma_c=1$ works as a boundary between the GL and non-GL type photon statistics. 

\subsection{E. Stochastic center manifold reduction}
Finally, we comment on the effect of the noises, in particular the carrier noise, on the center manifold reduction. In the previous subsection, we showed that the photon and carrier rate equations may be reduced to a single equation for photons around the lasing threshold. In that argument, we discussed only the deterministic rate equations and neglected the noise terms. According to the stochastic center manifold reduction developed in Refs \cite{Xu1996,Roberts2008}, noise terms originating from the carrier noise $f_N$ are added to the reduced equations of motion as a series of perturbations. Although the original stochastic rate equations (\ref{eq:REx})–(\ref{eq:REN}) include solely additive noises $f_{\alpha}$ and $f_N$, the reduced equations of motion can have multiplicative noises. Importantly, for the validity of the GL theory, the contribution of multiplicative noises must be negligible in the reduced photon equation of motion as in Eq. (\ref{eq:trans_normal}). Otherwise, multiplicative noises will make the steady-state distribution deviate from the GL distribution Eq. (\ref{eq:P_GL}) \cite{Stratonovich1967,Young1988,Fox1987}. In fact, with the numerical stochastic center manifold reduction \cite{Roberts}, we found that, in the reduced equation of motion, the lowest-order contribution from the carrier noise is $\beta\delta If_N$, which is a multiplicative noise term. Namely, the reduced equation of motion with the noise term is given as
\begin{eqnarray}
\delta \dot{I}&=&\gamma_c\delta\epsilon\delta I-\beta\gamma_c\delta I^2+f_I+\beta\delta If_N+\mathcal{O}(C(\delta I,\delta\epsilon))\nonumber\\
&&+({\rm higher}\ {\rm order}\ {\rm noise}\ {\rm terms}).
\end{eqnarray}
However, fortunately, since we assume that $\beta$ is much smaller than unity ($\beta\ll1$), even if the intensity (field) and carrier noise strength are comparable, the additive intensity (field) noise $f_I$ ($f_\alpha$) is expected to be dominant over the multiplicative noise originating from the carrier noise. Moreover, the higher-order noise terms are also negligible because they are proportional to $\beta^q$, where $q(\geq2)$ represents the order of a noise term.

\section{6. Discussion}
First, we comment on the laser design principle implied by our results. Our results indicate that, for low-$\beta$ lasers, effort to increase the $Q$ value is an important direction in terms of Poissonian light emission with low pump power for laser optical communications. Importantly, the conventional class-A condition ($\gamma_c\ll\gamma_\|$) is not necessary, and a photon lifetime comparable to the carrier lifetime ($\gamma_c\simeq\gamma_\|$) is sufficient to obtain a sharp drop in $g^{(2)}$ from 2 to 1 at the lasing threshold. Since carrier lifetimes in semiconductor lasers are on the order of nanoseconds, the required cavity photon lifetimes are also on the order of nanoseconds, which is realized with high-$Q$ cavities such as high-$Q$ photonic crystal cavities. 

On the other hand, extreme class-B lasers with $\gamma_c\ll\gamma_\|$ are also useful, for instance, as the light sources of two-photon excitation microscopies \cite{Jechow2013}, where bright thermal [$g^{(2)}>1$] light is required. When the photon lifetime is much shorter than the carrier lifetime, the thermal statistics remain far above the lasing threshold, which can be used  as a two-photon source with high intensity. Note that this condition is easily satisfied with low-$Q$ semiconductor lasers.

Experimentally, the measurement of higher-order photon correlations is available with novel techniques reported, for example, in Refs. \cite{Wiersig2009,Stevens2010,Elvira2011,Schlottmann2018}. However, obtaining the precise $\ln g^{(q)}$ vs. $\ln g^{(2)}$ shown in Fig. \ref{fig:classAB} is still too demanding. Therefore, in Appendix C, we propose a possible experimental method to check whether or not a given laser is described by the GL theory, which employs only light output intensity and $g^{(2)}$ measurements. We also note that the variation of the $Q$ value is reported in Ref. \cite{Baili2009}, which employed external cavities. 

\section{7. Conclusions}
First, we proposed a higher-photon correlation measurement method to confirm whether or not a laser is described by the Ginzburg-Landau (GL) theory. This technique allows the comparison of measured photon statistics with the GL theory without a photon (intensity) distribution function. Furthermore, in terms of experiments, this method has a great advantage in that the higher photon correlation functions are independent of quantum efficiencies.
 
Second, for low-$\beta$ lasers, we investigated the applicability of the GL theory for lasers with various photon and carrier lifetime ratios. When the photon lifetime is much longer than the carrier lifetime (class-A lasers), the photon statistics are described by the GL theory, which is easily understood in terms of conventional adiabatic elimination. Meanwhile, when the photon lifetime is much shorter than the carrier lifetime (class-B lasers), the photon statistics cannot be described by the GL theory. The surprise is the intermediate region. We found that even when the photon and carrier lifetimes are the same, the photon statistics are fully described by the GL theory. To interpret these results, using the center manifold reduction theory, which is an extension of adiabatic elimination, we showed that the GL theory is applicable even if the photon lifetime is equal to or longer than the carrier lifetime. In fact, the fundamental origin of the failure of the GL theory was found to be the photon-carrier relaxation oscillation. Thus, the applicability of the GL theory of lasers is broader than conventionally imagined. The implication of this conclusion reaches beyond theoretical interest and is important for laser design. 

\section{Acknowledgements}
We greatly appreciate Prof. A. J. Roberts for informing us of his stochastic center manifold reduction theory.

\renewcommand{\thefigure}{A-\arabic{figure}}
\setcounter{figure}{0}
\renewcommand{\theequation}{A-\arabic{equation}}
\setcounter{equation}{0}

\appendix
\section{Appendix A. Derivation of the GL potential through the Fokker-Planck equation}
In this appendix, following Refs. \cite{Risken1967,Haken2012,Risken1996,Lax1967,Louisell1973}, we derive the GL potential as a steady-state solution of the Fokker-Planck equation of lasers. We start from the normal form of the Hopf bifurcation with a noise term [Eq. (\ref{eq:Hopf})]
\begin{eqnarray}
\dot{\alpha}=\frac{1}{2}\mu\alpha-\frac{1}{2}\lambda|\alpha|^2\alpha+f_{\alpha},
\label{eq:Hopf_app}
\end{eqnarray}
where $\alpha$ is a complex value given by $\alpha=x+iy$, and the noise $f_\alpha=f_x+if_y$ is the Langevin noise.
The noise terms $f_x$ and $f_y$ satisfy the same correlations as in Eq. (\ref{eq:SL_noise}) in the main text, where the noise strength is represented by $Q$.
Now, the two-dimensional Fokker-Planck equation corresponding to Eq. (\ref{eq:Hopf_app}) is given by \cite{Risken1967}
\begin{eqnarray}
\frac{\partial P(x,y,t)}{\partial t}&=&\left[-\frac{\partial}{\partial x}\left\lbrace\frac{1}{2}\mu x-\frac{1}{2}\lambda(x^2+y^2)x\right\rbrace\right.\nonumber\\
&&-\frac{\partial}{\partial y}\left\lbrace\frac{1}{2}\mu x-\frac{1}{2}\lambda(x^2+y^2)y\right\rbrace\nonumber\\
&&\left.+\frac{\partial^2}{\partial x^2}\frac{Q}{2}+\frac{\partial^2}{\partial y^2}\frac{Q}{2}\right]P(x,y,t),
\label{eq:FPxy}
\end{eqnarray}
where $P(x,y,t)$ represents the probability distribution and can also be interpreted as the Glauber-Sudarshan P representation \cite{Walls2007}. In the polar coordinate defined as $\alpha=x+iy=re^{i\phi}$ (note that $r^2=I$ holds), the Fokker-Planck equation (\ref{eq:FPxy}) is rewritten as
\begin{eqnarray}
\frac{\partial P(r,\phi,t)}{\partial t}&=&\left[-\frac{1}{r}\frac{\partial}{\partial r}(\frac{\mu}{2} r^2-\frac{\lambda}{2} r^4)\right.\nonumber\\
&&\left.+\frac{Q}{2}\left\lbrace\frac{1}{r}\frac{\partial}{\partial r}\left(r\frac{\partial}{\partial r}\right)+\frac{1}{r^2}\frac{\partial^2}{\partial \phi^2}\right\rbrace\right]P(r,\phi,t).\nonumber\\
\label{eq:FP_polar}
\end{eqnarray}
Since the steady state must not have a preferred phase [$U(1)$ gauge symmetry], when the $\phi$ dependence is neglected, by setting $\dot{P}=0$, the steady-state $P_{\rm st}(\alpha)$ satisfies the following equation:
\begin{equation}
\frac{\partial}{\partial r}P_{\rm st}(r)=\frac{\frac{1}{2}\mu r-\frac{1}{2}\lambda r^3}{\frac{1}{2}Q}P_{\rm st}(r).
\end{equation}
Now, the steady-state solution is easily obtained as
\begin{equation}
P_{\rm st}(\alpha)=\frac{1}{Z}e^{-F(\alpha)},
\label{eq:Pst_app}
\end{equation}
where the potential $F(\alpha)$ is given by
\begin{eqnarray}
F(\alpha)&=&\frac{1}{\frac{1}{2}Q}\left(-\frac{1}{4}\mu r^2+\frac{1}{8}\lambda r^4\right)\nonumber\\
&=&\frac{1}{\frac{1}{2}Q}\left[-\frac{1}{4}\mu (x^2+y^2)+\frac{1}{8}\lambda(x^2+y^2)^2\right],
\end{eqnarray}
This is nothing else but the GL potential [see Eq. (\ref{eq:P_GL}) in the main text]. We note that the steady-state probability distribution given by Eq. (\ref{eq:Pst_app}) satisfies the detailed balance condition in the rotating frame of the laser frequency \cite{Risken1996}. Intuitively, this means that, in the rotating frame, there is no probability current in the steady-state solution $P_{\rm st}(\alpha)$.

\renewcommand{\thefigure}{B-\arabic{figure}}
\setcounter{figure}{0}
\renewcommand{\theequation}{B-\arabic{equation}}
\setcounter{equation}{0}

\section{Appendix B. Higher-order photon correlations and photon counting distribution function}
We comment on the relationship between the higher-order photon correlation $g^{(q)}$ and the photon counting distribution function $p_n$. We attempt to reconstruct the Glauber-Sudarshan P representation $P(I)$ ($I=|\alpha|^2$) and $p_n$ from $g^{(q)}$. Note that, here, $G^{(q)}$ denotes $G^{(q)}_\eta$ with $\eta=1$. First, with the Fourier transformation of $P(I)$, we introduce the characteristic function as \cite{Klauder2006,Risken1996}
\begin{eqnarray}
\Phi(t)&=&\int_{-\infty}^\infty P(I)e^{iIt}dI=\sum_{q=0}^{\infty}\frac{(it)^q}{q!}M_q.
\label{eq:charact_FT}
\end{eqnarray}
Here, the expansion coefficient $M_q$ is the $q$th moment and is equivalent to the non-normalized $q$th order moment $G^{(q)}$ as
\begin{eqnarray}
M_q=\int_{-\infty}^\infty I^qP(I)=\int_{0}^\infty I^qP(I)=G^{(q)},
\end{eqnarray}
where we used $P(I)=0$ for $I<0$. Recalling $g^{(q)}=G^{(q)}/\langle I\rangle^q$ with $\langle I\rangle=G^{(1)}$, the characteristic function is written as
\begin{eqnarray}
\Phi(t)=\sum_{q=0}^{\infty}\frac{\left(it\langle I\rangle\right)^q}{q!}g^{(q)}.
\label{eq:charact_gq}
\end{eqnarray}
Therefore, if infinite orders of the correlation $g^{(q)}$ are known, the ``shape" of the characteristic function can be calculated with Eq. (\ref{eq:charact_gq}). The mean photon number $\langle I\rangle$ works as a scaling factor for the characteristic function. With the characteristic function, the Glauber-Sudarshan P representation is calculated through the inverse Fourier transformation of Eq. (\ref{eq:charact_FT}) as
\begin{eqnarray}
P(I)=\frac{1}{2\pi}\int_{-\infty}^\infty \Phi(t) e^{-iIt}dt.
\end{eqnarray}
Finally, the photon counting distribution $p_n$ is obtained as
\begin{eqnarray}
p_n=\int_{0}^\infty dI \frac{I^n}{n!}e^{-I}P(I),
\label{eq:pn_P}
\end{eqnarray}
which is Eq. (\ref{eq:pn_eta}) for $\eta=1$. Now, we consider two concrete examples, namely far below and above the threshold of lasers described by the GL theory [see Eq. (\ref{eq:GL_gq})].

(i)  Far below the lasing threshold: we expect $g^{(q)}=q!$ from Eq. (\ref{eq:GL_gq}) far below the lasing threshold. Thus, the characteristic function will lead to $\Phi(t)=({1-it\langle I\rangle})^{-1}$. With the inverse Fourier transformation, we obtain the P representation of the exponential distribution: $P(I)=\langle I\rangle^{-1}\exp(-I/\langle I\rangle)$. Finally, with Eq. (\ref{eq:pn_P}), we obtain the photon counting distribution $p_n$ as
\begin{eqnarray}
p_n&=&\frac{1}{n!\langle I\rangle}\int_{0}^\infty dI I^ne^{-(1/\langle I\rangle+1)I}\nonumber\\
&=&\frac{\langle I\rangle^n}{(\langle I\rangle+1)^{n+1}}\ \ {\rm (far\ below\ threshold)},
\end{eqnarray}
which is the well-known thermal photon distribution, also called the Bose-Einstein distribution.

(ii)  Far above the lasing threshold: from Eq. (\ref{eq:GL_gq}), we expect $g^{(q)}=1$ for all $q$. Now, the characteristic function is given by $\Phi(t)=e^{it\langle I\rangle}$ The corresponding P representation is the delta function: $P(I)=\delta(I-\langle I\rangle)$ Finally, the photon counting distribution is 
\begin{eqnarray}
p_n&=&\int_{0}^\infty dI \frac{I^n}{n!}e^{-I}\delta(I-\langle I\rangle)\nonumber\\
&=&\frac{\langle I\rangle^n}{n!}e^{-\langle I\rangle}\ \ {\rm (far\ above\ threshold)},
\end{eqnarray}
which is also the well-known Poissonian photon distribution.

Therefore, it is, in principle, possible to reconstruct both the P representation and photon counting statistics from infinite orders of $g^{(q)}$. In practice, from finite orders of $g^{(q)}$, we may reconstruct the Glauber P representation using the maximum entropy method \cite{Gulyak2018}.

\renewcommand{\thefigure}{C-\arabic{figure}}
\setcounter{figure}{0}
\renewcommand{\theequation}{C-\arabic{equation}}
\setcounter{equation}{0}

\section{Appendix C. Possible experiments}
In this appendix, we propose a possible experiment for checking whether or not the GL theory is applicable for a given laser. Although there are state-of-the-art techniques for measuring higher-order photon correlations $g^{(q\geq3)}$ \cite{Wiersig2009,Stevens2010,Elvira2011,Schlottmann2018}, the precise measurement of photon statistics is still limited to the second-order photon correlation [$g^{(2)}$]. Therefore, here, we propose an alternative method that requires only the photon number and $g^{(2)}$ measurements.
\begin{figure}
\includegraphics[width=0.35\textwidth]{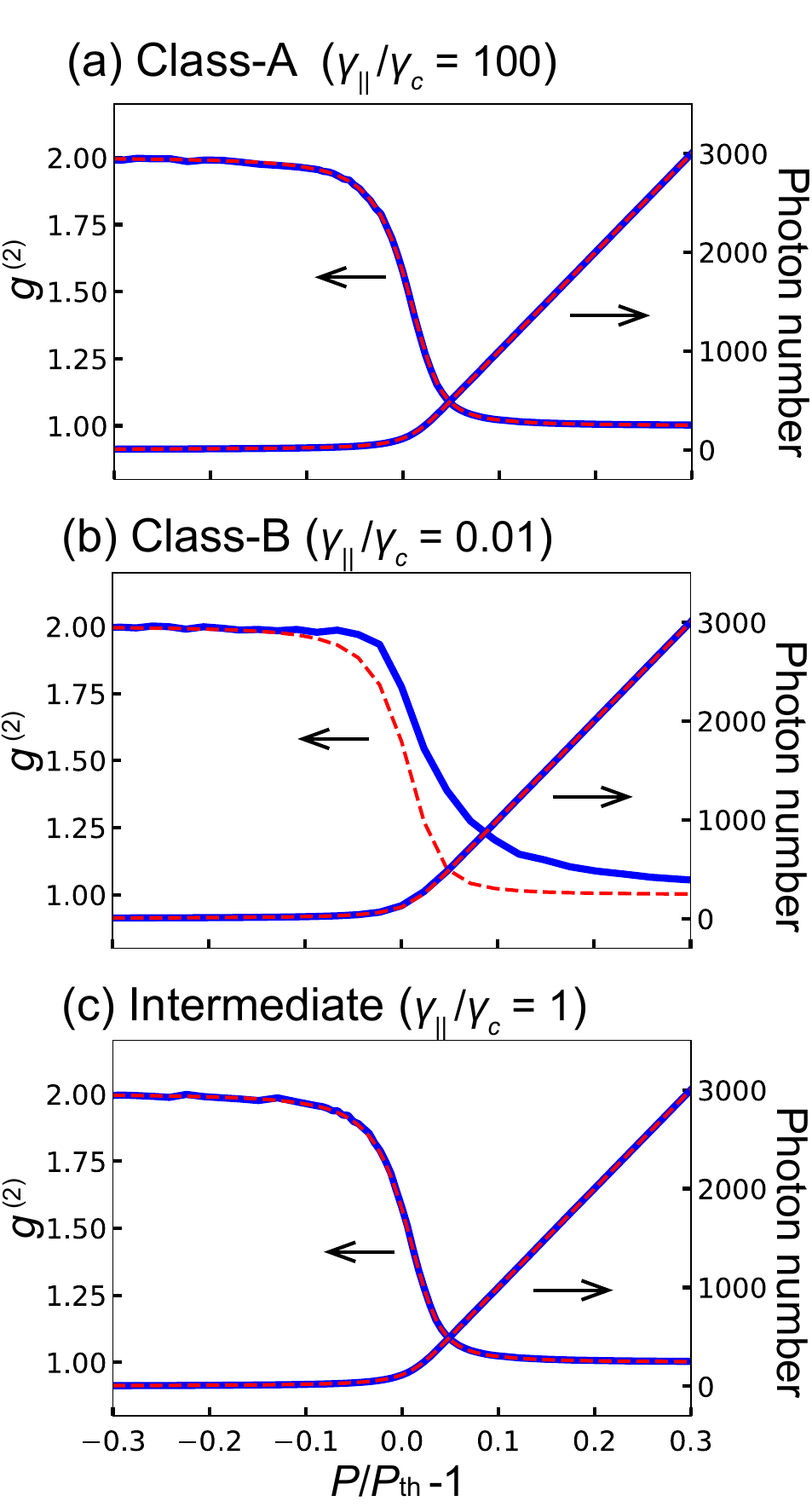}
\caption{(Color) Technique to check whether or not a given laser emission is described by the GL theory. The photon number and $g^{(2)}$ are represented by the blue solid curves, which are the same as those in the top row of Fig. \ref{fig:classAB}, but the photon number is shown on a linear scale. Meanwhile, the red dashed curves are the photon number and $g^{(2)}$ predicted with the GL theory [Eq. (\ref{eq:fitting})]. Here, (a), (b), and (c) are respectively for class-A ($\gamma_\|/\gamma_c=100$), class-B ($\gamma_\|/\gamma_c=0.01$), and intermediate ratio ($\gamma_\|/\gamma_c=1$).For (a), (b), and (c), we used $\beta=10^{-4}$.}
\label{fig:fitting}
\end{figure}

The strategy is to make use of the fact that, in the GL theory, the photon output, which is actually $G^{(1)}$, and $g^{(2)}$ are not independent but instead correlated as Eqs. (\ref{eq:unnorm_Gq}) and (\ref{eq:GL_gq}). First, by measuring a pump-input and light-output curve, we fit it with a function 
\begin{equation}
y(x)=A\frac{D_{-2}(Bx)}{D_{-1}(Bx)},
\label{eq:fitting}
\end{equation}
where $A$ and $B$ are fitting parameters. Although light output intensity strongly depends on the generalized quantum efficiency of a detector, $\eta$, the shape of a pump-input and light-output curve can generally be fitted with Eq. (\ref{eq:fitting}). Second, with the fitting parameter $B$ obtained from the first fitting, we plot the $g^{(2)}$ as
\begin{eqnarray}
g^{(2)}=\frac{2D_{-3}(Bx)D_{-1}(Bx)}{[D_{-2}(Bx)]^2},\nonumber\\
\label{eq:fitting_g2}
\end{eqnarray}
which is the prediction of the GL [see Eq. (\ref{eq:GL_gq})]. In Fig. \ref{fig:fitting}, we show the simulated pump-input and light output curves and $g^{(2)}$ on linear scales with blue solid curves. Meanwhile, the red dashed curves in Fig. \ref{fig:fitting} represent the fitting curves of the pump-input and light outputs [Eq. (\ref{eq:fitting})] and theoretical prediction of [Eq. (\ref{eq:fitting_g2})]. In Fig. \ref{fig:fitting}, we used $\beta=10^{-4}$, while the curves in (a), (b), and (c)  are for class-A ($\gamma_\|/\gamma_c=100$), class-B ($\gamma_\|/\gamma_c=0.01$), and intermediate ratio ($\gamma_\|/\gamma_c=1$), respectively. Therefore, the simulated pump-input and light output curves and $g^{(2)}$ represented by the blue curves are the same as those in the top row of Fig. \ref{fig:classAB}. Since the pump-input and light output curves, which are determined only by $\beta$, are the same for (a), (b), and (c), they can be well fitted with the same values of $A$ and $B$. On the other hand, for $g^{(2)}$, the predictions of the GL theory well fit $g^{(2)}$ in (a) and (b), while the GL theory cannot fit $g^{(2)}$ in (b). This result can be expected from the discussion in the main text. Therefore, Fig. \ref{fig:fitting} indicates that the validity of the GL theory for a given laser can be checked solely with a measured pump-input and light-output curve and $g^{(2)}$.

\renewcommand{\thefigure}{D-\arabic{figure}}
\setcounter{figure}{0}
\renewcommand{\theequation}{D-\arabic{equation}}
\setcounter{equation}{0}

\section{Appendix D. Toda oscillator approach to class-B lasers}
Here, we briefly discuss the pioneering investigations of class-B lasers based on the nonlinear (Toda) oscillator model \cite{Oppo1985,Paoli1988,Lien2002}. Although a full understanding of the connection between our center manifold reduction approach and the Toda oscillator approach is far beyond the scope of this paper, it is still interesting to review some insights provided by our approach into these previous studies on class-B lasers.  

It was proved by Oppo and Politi  \cite{Oppo1985} that dynamics represented by the rate equations (\ref{eq:REp_MF}) and (\ref{eq:REn_MF}) are equivalent to Hamiltonian dynamics in a Toda-like potential with ``position"-dependent friction. In this framework, for $\gamma_\|/\gamma_c\ll1$,  the system exhibits phase spatial periodic rotation, while the ``position"-dependent friction gives rise to the contraction of the phase space, which is nothing else but the photon-carrier relaxation oscillation of class-B lasers discussed in the main text. Ref. \cite{Oppo1985} also pointed out that, when $\gamma_\|/\gamma_c\gg1$, dynamics in the Toda-like potential are reduced to the motion in the GL potential described as Eq. (\ref{eq:van_der_Pol}). Thus, the framework developed in Ref. \cite{Oppo1985} would be an alternative approach to understand the mechanism of the feasibility and breakdown of the GL theory for lasers with $\gamma_\|/\gamma_c\simeq1$ from the standpoint of a Hamiltonian system \cite{Graham1984}.  
\begin{figure}
\includegraphics[width=0.39\textwidth]{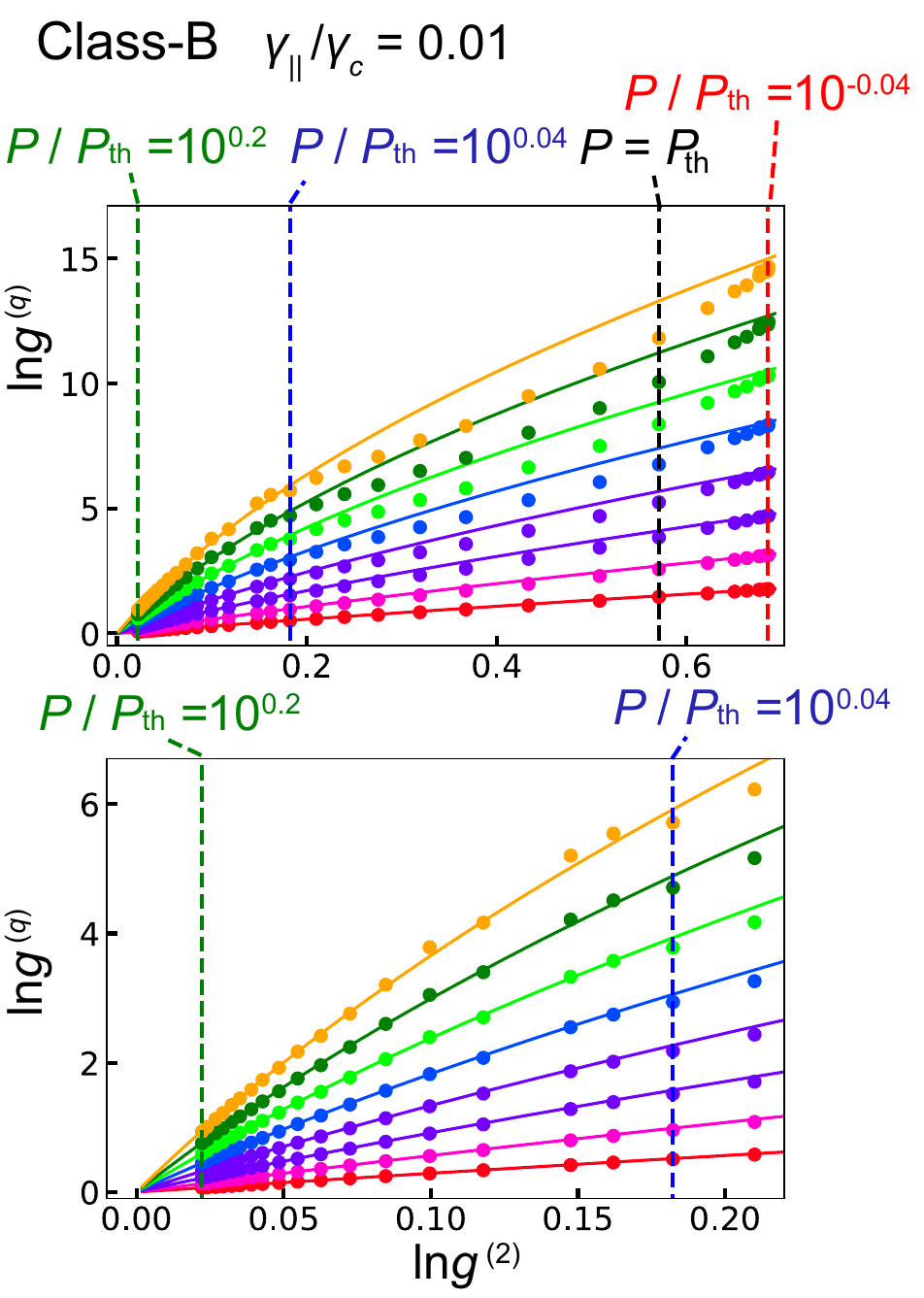}
\caption{(Color) The filled colored circles represent simulated $\ln(g^{(q)})$ vs. $\ln(g^{(2)})$ for class-B lasers ($\gamma_c/\gamma_\|=100$) with $\beta=10^{-4}$, which are the same as those in the bottom row of Fig. \ref{fig:classAB}(b).
The solid colored curves are analytically obtained $\ln(g^{(q)})$ vs. $\ln(g^{(2)})$ based on the intensity distribution $P^{\rm Toda}(\alpha)$ [Eq. (\ref{eq:Toda_alpha})]. The lower row is the zoomed-in $\ln(g^{(q)})$ vs. $\ln(g^{(2)})$ of the upper row. The vertical dashed lines represent specific pump powers, $P/P_{\rm th}=10^{-0.04}$, 1, $10^{0.04}$, and $10^{0.2}$.}
\label{fig:Toda_scaling}
\end{figure}

Furthermore, the Fokker-Planck equation corresponding to the stochastic Toda oscillator for class-B lasers was investigated by Paoli, Politi, and Arrechi \cite{Paoli1988}, which derived that in a certain parameter regime, the intensity distribution (the solution of the Fokker-Planck equation) can be approximately given in the form
\begin{equation}
P^{\rm Toda}(\alpha)=Z^{-1}I^Ae^{-BI}\ \ \ A\geq0\ {\rm and}\ B>0,
\label{eq:Toda_alpha}
\end{equation}
with $I=|\alpha|^2$ and $Z=A^{-(B+1)}\Gamma(B+1)$. 
First, it should be noted that $P^{\rm Toda}(\alpha)$ is \textit{not} the potential solution as expressed in Eq. (\ref{Pfunction}).
Second, note that, interestingly, Ogawa also derived the intensity distribution of the form Eq. (\ref{eq:Toda_alpha}) \cite{Ogawa1989,Ogawa1990} but for the bad-cavity limit, where the field degree of freedom was adiabatically eliminated, and thus is not exactly the same as a class-B laser considered here.

Now, by using Eq. (\ref{eq:Gq_general}), the $q$th order moment $G^{(q)}_\eta$ corresponding to $P^{\rm Toda}(\alpha)$ is calculated as
\begin{eqnarray}
G^{(q)}_\eta=\eta^q\int_0^\infty dI\ I^qP(I)=\eta^qB^{-q}\frac{\Gamma(q+A+1)}{\Gamma(A+1)}.\nonumber\\
\label{eq:Toda_moment}
\end{eqnarray}
The normalized $q$th order normalized photon correlation function $g^{(q)}$ is given by
\begin{eqnarray}
g^{(q)}=\frac{G^{(q)}_\eta}{(G^{(1)}_\eta)^q}=\frac{\Gamma(A+q+1)\Gamma(A+1)^{q-1}}{\Gamma(A+2)^q}.
\label{eq:gq_Toda}
\end{eqnarray}
Interestingly, the parameter $B$ does not appear in the normalized higher-order photon correlation $g^{(q)}$. This is because Eq. (\ref{eq:Toda_alpha}) can always be rewritten as $P^{\rm Toda}(\alpha)={Z^\prime}^{-1} {I^\prime}^A\exp(-I^\prime)$ by introducing new parameters $Z^\prime=ZB^A$ and $I^\prime=BI$. This means that the parameter $B$ does not affect the ``shape" of the intensity distribution. Therefore, as with GL-type photon statistics, the plot $\ln(g^{(q)})$ vs. $\ln(g^{(2)})$ is again a powerful tool to probe whether or not photon statistics can be described by Eq. (\ref{eq:Toda_alpha}) without using any fitting parameter.
Before comparing  $\ln(g^{(q)})$ vs. $\ln(g^{(2)})$ with numerical simulation, let us discuss $g^{(2)}$ and $G^{(1)}(=\langle I\rangle)$, which are calculated respectively as
\begin{eqnarray}
G^{(1)}=\langle I\rangle=B^{-1}(A+1)
\label{eq:G1_Toda}
\end{eqnarray}
and
\begin{eqnarray}
g^{(2)}=1+\frac{1}{A+1}.
\label{eq:g2_Toda}
\end{eqnarray}
According to Eqs. (\ref{eq:G1_Toda}) and (\ref{eq:g2_Toda}), when the parameter $A$ is varied from 0 to $+\infty$, $\langle I\rangle$ linearly increases and $g^{(2)}$ changes from $g^{(2)}=2$ to 1. Therefore, the parameter $A$ seems to work as a pump-like parameter.

Now, in Fig. \ref{fig:Toda_scaling}, the analytically obtained $\ln g^{(q)}$ vs. $\ln g^{(2)}$ [see Eq. (\ref{eq:gq_Toda})] are represented by solid colored curves. Meanwhile, the filled colored circles in Fig. \ref{fig:Toda_scaling} represent numerically simulated $\ln g^{(q)}$ vs. $\ln g^{(2)}$ for the class-B laser ($\gamma_|/\gamma_c=0.01$) with $\beta=10^{-4}$, which are the same data as in the bottom row in Fig. \ref{fig:classAB}(b).
Figure \ref{fig:Toda_scaling} indicates that, around the lasing threshold, the filled circles clearly deviate from the solid curves (see the region from $P/P_{\rm th}=10^{-0.04}$ to $10^{0.04}$), but they start to fall on the solid curves well above the lasing threshold (see the region above $P/P_{\rm th}=10^{0.04}$), which is more evident for the zoomed-in $\ln g^{(q)}$ vs. $\ln g^{(2)}$ shown in the lower row of Fig. \ref{fig:Toda_scaling}. Therefore, we conclude that the photon statistics of the class-B laser asymptotically coincides with the intensity distribution $P^{\rm Toda}(\alpha)$ [see Eq. (\ref{eq:Toda_alpha})] with an increase in pump power. 

Interestingly, when pump power is high enough, the photon statistics of class-B lasers seems to be described by the intensity distribution $P^{\rm Toda}(\alpha)$ independently of the detailed values of the parameters $\gamma_|/\gamma_c$ and $\beta$.
As discussed in the main text, in the lasing threshold region, lasers with $\gamma_|/\gamma_c\geq1$ have universality in that their dynamics can be described by GL theory, while class-B lasers $\gamma_|/\gamma_c\geq1$ do not. In fact,  around the lasing threshold, the ratio $\gamma_|/\gamma_c\geq1$ strongly affects the dynamics of a class-B laser, for instance, the frequency and damping rate of photon-carrier relaxation oscillation [see Eq. (\ref{eq:eigen_below})]. Therefore the fact that even class-B lasers seems to have the universality in the high pump power regime is very surprising. The origins of the intensity distribution $P^{\rm Toda}(\alpha)$ and its universality are not yet clear, and gaining an intuitive understanding of them is important future work.

%

\end{document}